\documentclass[printer]{aa}  

\usepackage{graphicx}
\usepackage{txfonts}
\usepackage{color}
\usepackage{xcolor}
\usepackage{amsmath}
\usepackage[]{hyperref}
\begin{document} 

   \title{XMM/HST monitoring of the ultra-soft highly accreting Narrow Line Seyfert 1 RBS\,1332}

%   \subtitle{Dissertation on a pretty little dog}

   \author{R. Middei 
          \inst{1,2}\fnmsep\thanks{riccardo.middei@ssdc.asi.it}
          \and
           S. Barnier \inst{3},
           F. G. Saturni \inst{1,2}
           \and F. Ursini \inst{5}
           \and P.-O. Petrucci \inst{4}
           \and S. Bianchi \inst{5}
            \and M. Cappi \inst{6}
            \and M. Clavel \inst{4}
            \and B. De Marco \inst{7}
            \and A. De Rosa \inst{8}
            \and G. Matt \inst{5}
           \and G. A. Matzeu \inst{9}
           \and M. Perri \inst{1,2}
          }

   \institute{ 
   INAF Osservatorio Astronomico di Roma, Via Frascati 33, 00078 Monte Porzio Catone (RM), Italy.         
   \and   Space Science Data Center, Agenzia Spaziale Italiana, Via del Politecnico snc, 00133 Roma, Italy.      
   \and Department of Earth and Space Science, Graduate School of Science, Osaka University, Toyonaka, Osaka 560-0043, Japan.
   \and Univ. Grenoble Alpes, CNRS, IPAG, 38000 Grenoble, France.
   \and Dipartimento di Matematica e Fisica, Universit\`a degli Studi Roma Tre, Via della Vasca Navale 84, 00146 Roma, Italy.
   \and INAF-Osservatorio di Astrofisica e Scienza dello Spazio di Bologna, Via Gobetti, 93/3, 40129 Bologna, Italy.
   \and Departament de Fìsica, EEBE, Universitat Polit\'ecnica de Catalunya, Av. Eduard Maristany 16, 08019 Barcelona, Spain.
   \and INAF, Istituto di Astrofisica e Planetologia Spaziali, Via Fosso del Cavaliere, 100 - I-00133 Rome, Italy.
   \and Quasar Science Resources SL for ESA, European Space Astronomy Centre (ESAC), Science Operations Department, 28692, Villanueva de la Ca\~{n}ada, Madrid, Spain.
   }
  %    
    %     
       %  

            % \and LaLaland
             %\and chiuahua
             %\and cavalierkingstore
             %\and dorgistore            
             %\and pugstore

   \date{Received mm/dd/yyyy; accepted mm/dd/yyyy}

% \abstract{}
% 5 {} token are mandatory
 
\abstract  
{Ultra-soft narrow line Seyfert 1 (US-NLSy) are a poorly observed class of active galactic nuclei characterized by significant flux changes and an extreme soft X-ray excess. This peculiar spectral shape represents a golden opportunity to test whether the standard framework commonly adopted for modelling local AGN is still valid. 
We thus present the results on the joint \textit{XMM-Newton} and 
\textit{HST} monitoring campaign of the highly accreting US-NLSy RBS\,1332. The optical-to-UV spectrum of RBS\,1332 exhibits evidence of both a stratified narrow-line region and an ionized outflow, that produces absorption troughs over a wide range of velocities (from $\sim$--1500 km s$^{-1}$ to $\sim$1700 km s$^{-1}$) in several high-ionization transitions (Ly$\alpha$, N {\footnotesize V}, C {\footnotesize IV}). From a spectroscopic point of view, the optical/UV/FUV/X-rays emission of this source is due to the superposition of three distinct components which are best modelled in the context of the two-coronae framework in which the radiation of RBS\,1332 can be ascribed to a standard outer disk, a warm Comptonization region and a soft coronal continuum. The present dataset is not compatible with a pure relativistic reflection scenario. Finally, the adoption of the novel model \textsc{reXcor} allowed us to determine that the soft X-ray excess in RBS\,1332 is dominated by the emission of the optically thick and warm Comptonizing medium, and only marginal contribution is expected from relativistic reflection from a lamppost-like corona.
}
%-- which is potentially contaminated by the emission of an interloper Wolf-Rayet star --


   \maketitle
%
%-------------------------------------------------------------------
%

\section{Introduction}

\indent Active galactic nuclei (AGN) are compact sources at the center of galaxies. They are powered by accretion of matter onto a supermassive black hole and emit across the whole electromagnetic domain, from radio up to $\gamma$-rays \citep[e.g.][]{Padovani2017}.\\
\indent AGN are characterized by a significant amount of energy emitted in the optical/UV and X-ray ranges. This emission is accounted for by the so-called two phase model \citep{Haardt1991,Haardt1993}. Seed optical/UV photons from an optically thick geometrically thin accretion disc are Compton scattered to the X-rays by a thermal distribution of electrons, the so-called hot corona. This mechanism explains the power-law shape of the X-ray spectrum and the observations of a high-energy cutoff at a few hundred keV in most sources \citep[][]{Perola2000,Malizia2014,Fabian2015,Tortosa2018,Kamraj2022}. The reprocessing of the primary X-ray continuum by the disc or more distant material gives rise to features like a Fe K$\alpha$ emission line \citep[e.g.][]{George1991}. Most objects also show the presence of a soft X-ray excess, below $\sim$2 keV, with respect to the high-energy power-law extrapolation \citep[e.g.][]{Bianchi2009,Gliozzi2020}.\\
\indent The physical origin of this soft X-ray excess component is still highly debated.
So far, two models have been commonly adopted to describe such a component: relativistic reflection and warm Comptonization. In the first case, it is assumed that the soft X-ray excess results from the blending of emission lines due to relativistic reflection \citep[e.g.][]{Crummy2006}. On the other hand, an alternative explanation invokes an additional emitting component, the so-called warm corona, which is described as an optically thick and geometrically thin warm plasma above the accretion disc \citep[e.g.][]{Magdziarz1995,Petrucci2013}. This component can be modeled as fully covering the accretion disc \citep[the disc in this case is assumed to be passive][]{Petrucci2018}, or as a patchy medium \citep[][]{Kubota2018}. partially covering the accretion disc.
From an observational X-ray perspective, the soft excess of AGN have been successfully reproduced using a pure relativistic reflection scenario \citep[][]{Walton2013,Mallick2018,Jiang2020,Xu2021}. The analysis of multiwavelength multi-epoch datasets proved to be a powerful tool for studying the origin of the soft X-ray excess. A broadband coverage enables us to model and disentangle the different spectral components and, through variability, test the connection among them \citep[][]{Middei2023eso,Mehdipour2023}. In these cases, the joint fit of the UV and X-ray data shows that warm Comptonization is a viable model to explain the origin of the soft X-ray excess and is generally statistically favored by the data compared to a pure relativistically blurred reflection model \citep[e.g.][]{Matzeu2020,Middei2020,Porquet2021,Porquet2024}.\\
\indent In the context of this approach, we complemented our previous spectral and variability campaigns with a new series of \textit{XMM-Newton} and \textit{HST} observations of RBS\,1332 \citep[z=0.122, ][]{Bade1995}. This source is classified as an Ultra Soft Narrow Line Seyfert 1 (US-NLSy 1), AGN potentially hosting BHs accreting around or above the Eddington limit, and with inner disc regions characterized by higher temperatures with respect to standard Seyfert galaxies. 
In particular, the disc emission is expected to peak in the Far Ultra Violet (FUV) rather than in the UV. They are also characterised by strong Fe II emission \citep[e.g.][]{Osterbrock1977,Goodrich1989}. In the X-rays, they exhibit a much steeper continuum in comparison with average Seyferts (i.e. $\Gamma$ in the range 2.0-2.5  instead  of 1.5-2, e.g. \citealt[][]{ Bianchi2009}), and a strong,  highly variable soft excess  \citep[][]{Gallo2018}. The origin of their broadband continuum is still debated. In fact, although \citet{Jiang2020} successfully modelled the \textit{XMM-Newton} data of five US-NLSys with relativistic blurred reflection model, for at least one of these sources (RX J0439.6-5311) the two-coronae model was found to provide also a good fit \citep[][]{Jin2017,Jin2017b}. 
%#
RBS\,1332 has very low intrinsic absorption \citep[$\rm E_{B-V}=0.008$][]{Grupe2010}, making it ideal for studying the relation between the soft X-ray band and UVs and \citet{Grupe2004} reported on the optical properties of this NLSy1 source with FWHM(H$_{\beta}$)=1100 km s$^{-1}$ and a H$\alpha$/H$\beta\sim$3.3.
%#
\indent In this paper we will thus test the origin of the soft X-ray excess in RBS\,1332 studying a broadband multi-epoch dataset taken with \textit{XMM-Newton} and \textit{HST}. The paper is organised as follows: The data reduction is discussed in Sect. 2; Sect.s 3 and 4 report on the UV and UV-to-X-ray spectral analysis, respectively. Finally, in Sect. 5 we discuss and comment on our findings.

\section{Data processing and reduction}
 \begin{figure*}
	\centering
	\includegraphics[width=\textwidth]{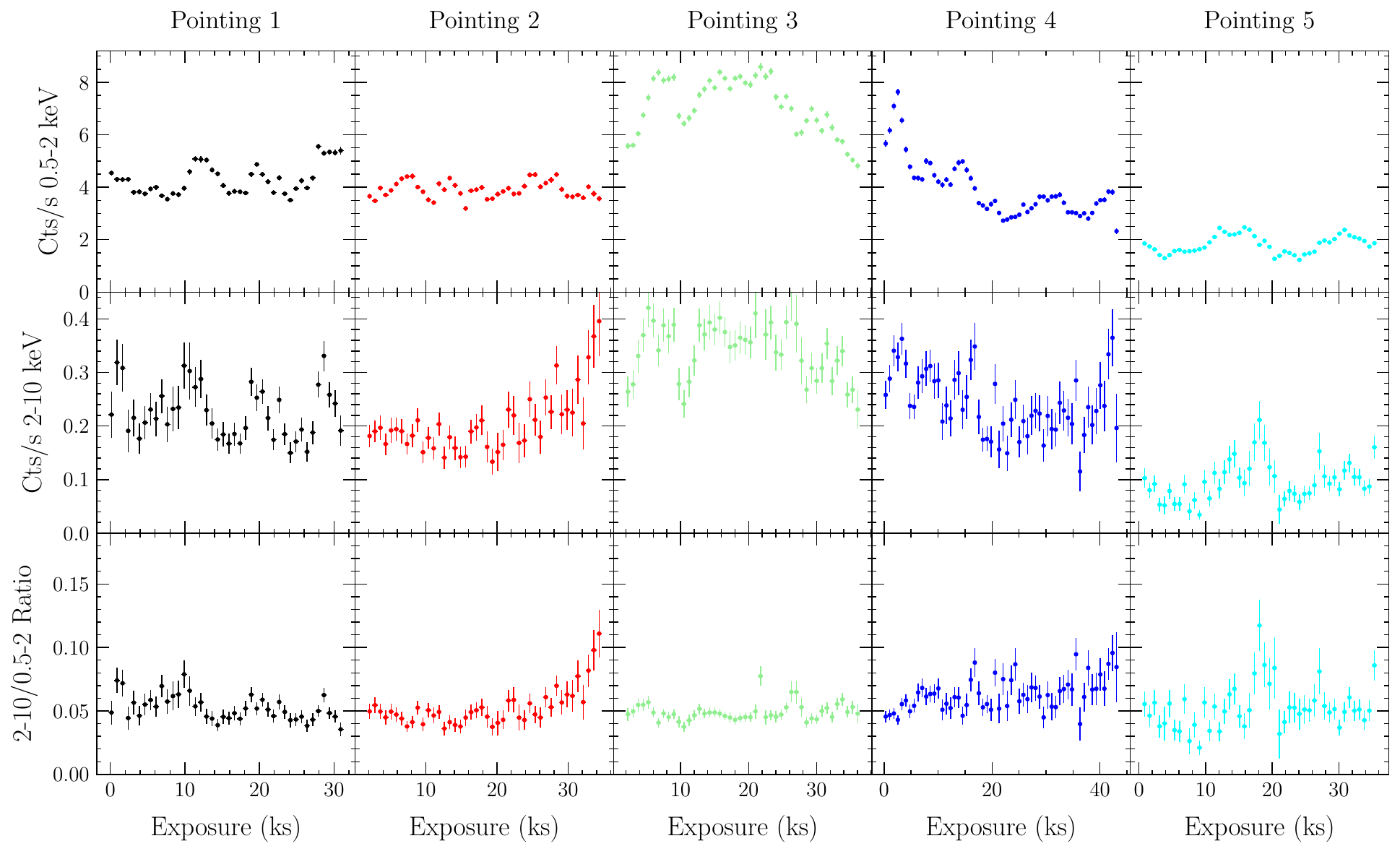}
	\caption{Multi-epoch X-ray time series of RBS\,1332. The soft X-rays show a significant variability compatible with flux changes observed above 2 keV. The ratios between hard and soft X-rays also exhibit changes on ks timescales, especially in observations 2 and 5, see the Sect. 2 for model details. Black, red, green, blue, magenta and cyan colours refer to obs. 1,  obs. 2, obs. 3, obs. 4 and obs. 5, respectively. This color code is adopted in the whole paper.}
	\label{lc}
\end{figure*}

The  RBS\,1332 data analysed here belongs to the joint \textit{XMM-Newton/HST} observational campaign consisting in 5 $\times$ (20 ks (\textit{XMM-Newton}) + 1 orbit (\textit{HST})) quasi-simultaneous observations. The exposures cover the time period between November 06$^{th}$ and 19$^{th}$ 2022, with consecutive pointings being about two or three days apart, see Table~\ref{log}. Unfortunately, due to star tracker errors, \textit{HST} was unable to observe during epochs 2 and 3.\\
\indent \textit{XMM-Newton} data of RBS\,1332 were obtained with  the EPIC cameras \citep[][]{Struder2001,Turner2001} in Small Window mode with the medium filter applied. Science products are obtained processing  the \textit{XMM-Newton} Science Analysis System (SAS, Version 21.0.0). The source extraction radius and the screening for high background time intervals were executed adopting an iterative process that maximizes the S/N (as described in  \citealt[][]{Piconcelli2004,Nardini2019}). The source radii span between 29 and 34 arcsec, while the background was computed from a blank region with radius 40 arcsec.  Spectra are later binned to have at least 30 counts in each bin, and not to oversample  the instrumental energy resolution by a factor larger  than 3. We also extract data provided by the Optical Monitor \citep[][]{Mason2001}, on-board \textit{XMM-Newton}. RBS\,1332 was observed with the filters UVW1 (2910 \AA), UVM2 (2310 \AA), UVW2 (2120 \AA) during the whole monitoring program. Data provided by the OM were extracted using the standard procedure within SAS and we converted the spectral points into an \textsc{XSPEC} \citep[][]{Arnaud1996} compliant format using the task \textsc{om2pha}.\\
\indent In Fig.~\ref{lc} we show the soft (0.3-2 keV) and hard (2-10 keV) light curves (top and middle panels) and their ratios (bottom panels). In accordance with this figure, flux variability of up to 30\% on hourly timescale is observed. On very short timescales (of a few ks),
the hardness ratios show hints of mild spectral variability, especially in the final part of pointing 2 and for about $\sim$6 ks in observation 5. However, the short duration of these events does not allow us to obtain a sufficiently high S/N for a detailed time-resolved analysis, therefore we decided to average over the entire exposures.

\indent RBS\,1332 was also observed by the Hubble Space Telescope using the Cosmic Origins Spectrograph (COS). This spectrograph \citep[][]{Green2012} enables to perform high-sensitivity, medium- and low-resolution (1500--18000) spectroscopy in the 815--3200 \AA\ wavelength interval. In this case, grating G140L centered at 1280 \AA\ was adopted, thus enabling us to measure the source emission between 1100 and 2280 \AA\ (observer frame).  Final spectra were obtained though the automatic \textit{HST}/COS calibration pipeline CalCOS.

\begin{figure}
	\centering
	\includegraphics[width=\columnwidth]{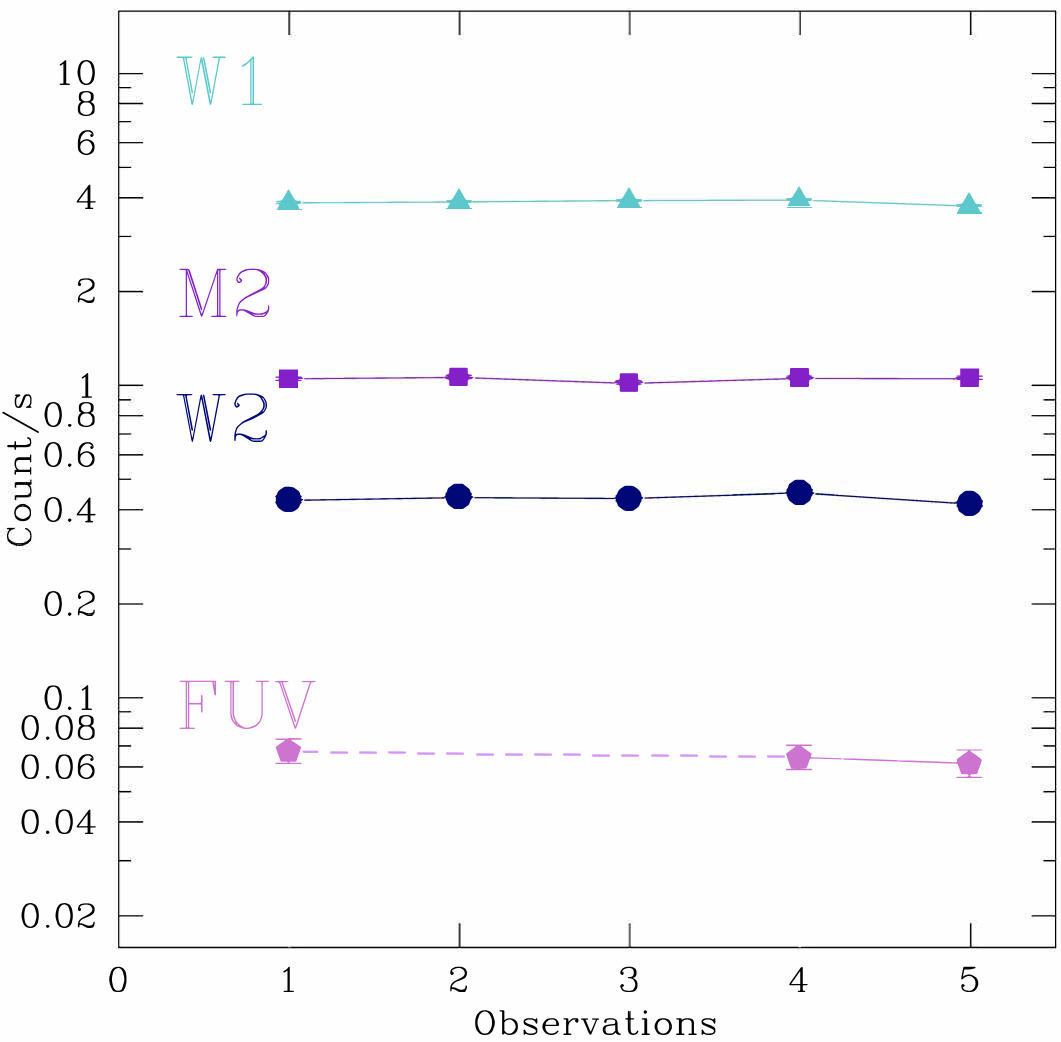}
	\caption{UV light-curves obtained using XMM-Newton optical monitor and HST (FUV).}
	\label{omlc}
\end{figure}

\indent In Fig.~\ref{omlc} we show the OM and COS rates as observed by \textit{XMM-Newton} and the \textit{HST} telescope. Differently to the X-rays flux, no significant flux evolution is observed in optical/UV during the observational campaign.

\begin{table}
		\centering
		\caption{\small{Log of the observations \textit{XMM-Newton}-\textit{HST} monitoring campaign.}\label{log}}
		\begin{tabular}{c c c c}
			\hline
			Observatory & Obs. ID & Start date & Net exp. \\
			& & yyyy-mm-dd & ks \\
			\hline 
			\textit{XMM-Newton} &0903370201  &2022-11-06  & $\sim$22\\
      		\textit{HST} &  &	
			2022-11-06 & $\sim$2 \\
            \hline
			\textit{XMM-Newton} &0903370301  &2022-11-09 & $\sim$23\\
			\hline
			\textit{XMM-Newton} &0903370401  & 2022-11-12  & $\sim$24\\
			\hline
			\textit{XMM-Newton} &0903370501  &	
			2022-11-15 & $\sim$30 \\
   		\textit{HST} &  &	
			2022-11-16 & $\sim$2 \\
			\hline
			\textit{XMM-Newton} &0903370601  &	
			2022-11-18  & $\sim$25 \\
   			\textit{HST} &  &	2022-11-19
			  & $\sim$2 \\
			\hline
		\end{tabular}
\end{table}

%Except for the 2015 MOS data, all the other exposures are some-
%#what spoiled by some background flares. These periods were filtered via an iterative process aimed at maximising the S/N ratio in the 4–9 keV energy band (e.g., Piconcelli et al. 2004; Nardini et al. 2019).

\section{HST analysis}\label{sec:hst-ana}
We started our investigation on the broadband properties of RBS\,1332 characterizing its \textit{HST} spectrum. First, we visually inspected the main UV emission lines in search of relevant absorption features. In doing this, we found that the Ly$\alpha$, the N {\footnotesize V} and the C {\footnotesize IV} transitions were affected by absorption. The position and shape of each trough are similar in all the emission features, with only the ``red'' Ly$\alpha$ absorption missing. This hints at the fact that such absorption features are produced by material intrinsic to RBS\,1332 that is located in the surrounding environment of the AGN central engine; to accurately quantify its properties, we performed a spectral decomposition of every absorbed emission system.

\begin{figure*}[htbp]
    \centering
    \includegraphics[width=\textwidth]{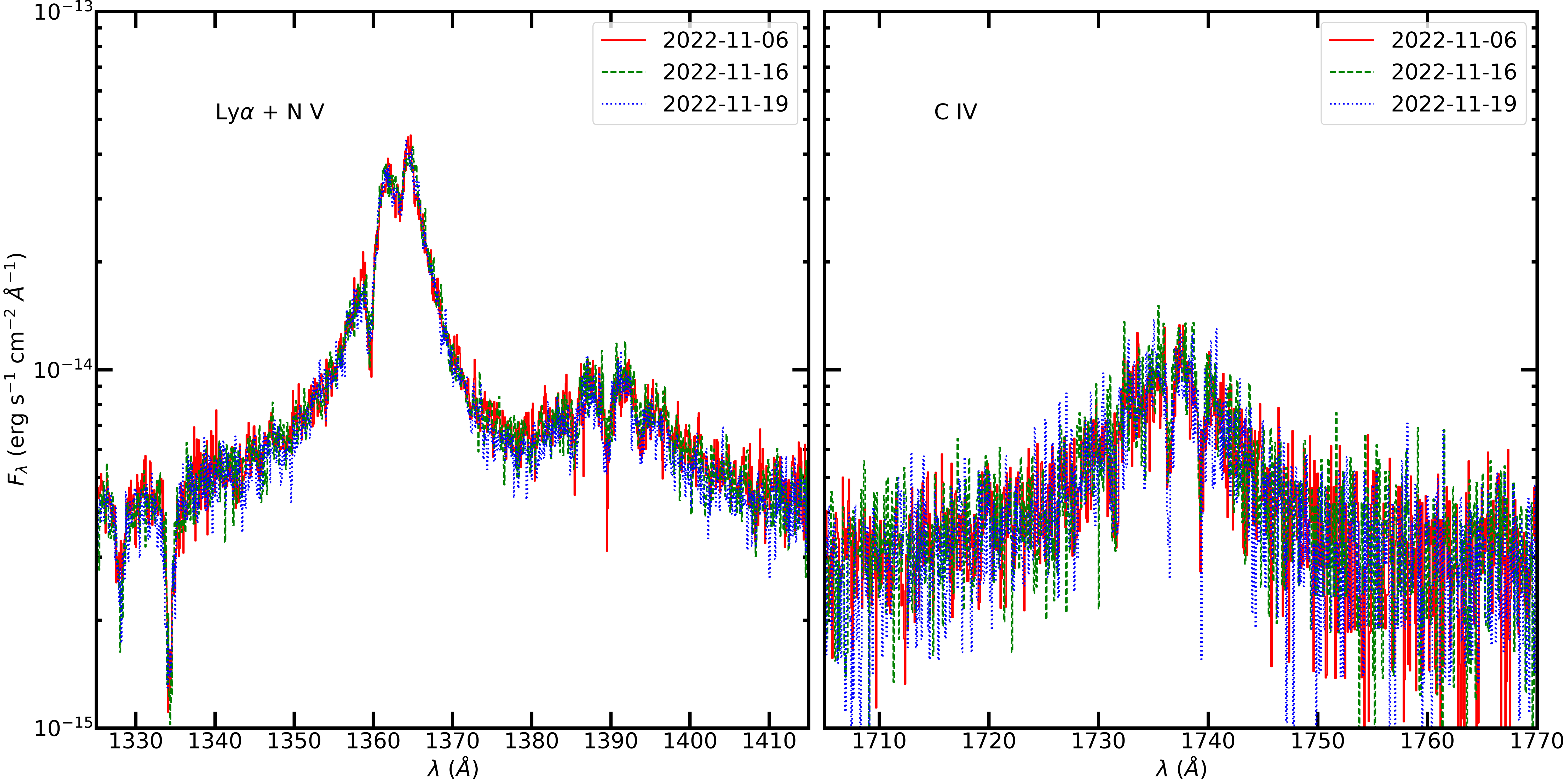}
    \caption{Comparison of the RBS\,1332 Ly$\alpha$+N {\scriptsize V} and C {\scriptsize IV} spectral regions in the observer frame over the three {\it HST} observation epochs ({\it see legend}). {\it Left panel:} Ly$\alpha$+N {\scriptsize V} spectral region. {\it Right panel:} C {\scriptsize IV} spectral region.}
    \label{fig:comp}
\end{figure*}

We show the Ly$\alpha$+N {\footnotesize V} and C {\footnotesize IV} spectral regions of each observation in Fig. \ref{fig:comp}: at a glance, all absorption features appear to be relatively stable in both shape and depth within the spectral noise over the three {\it HST} epochs (see Tab. \ref{log}). This implies that the absorber is neither changing in structure \citep[e.g.,][]{Krongold2010,Hall2011} nor responding to variations in the ionizing flux \citep[e.g.,][]{Barlow1992,Trevese2013} over the course of the {\it HST} observations; this allowed us to combine them together by averaging them into a single UV spectrum in the observer-frame interval 1110--2280 \AA\, (corresponding to an interval 990--2030~\AA\, in the rest frame). On this spectrum, we identified intervals that are relatively free of major emission and absorption features, and used them to compute a power-law continuum  $F_\lambda$ of the form:
\begin{equation}
    F_\lambda \propto \lambda^\alpha
\end{equation}
with $\alpha$ the spectral index. The result fits well the RBS\,1332 continuous emission over the entire wavelength interval, with $\alpha = -1.37 \pm 0.10$; therefore, we adopt this fit as a good representation of the actual RBS\,1332 continuum in the following analysis of the AGN emission and absorption lines.

\begin{figure}[htbp]
    \centering
    \includegraphics[width=1.1\columnwidth]{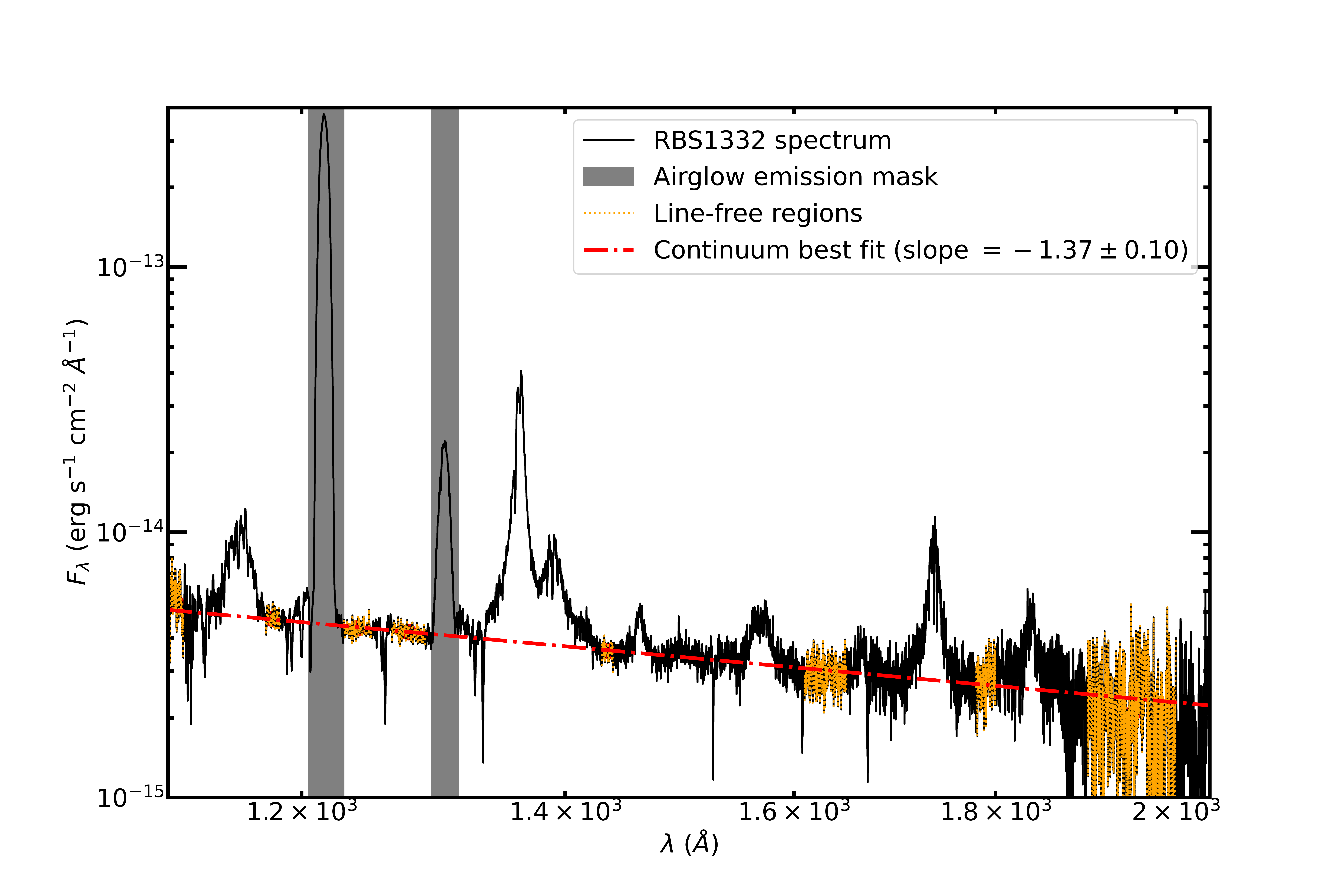}
    \caption{HST spectrum of RBS\,1332 ({\it black solid line}). The continuum emission ({\it red dot-dashed line}) fitted over intervals free of major emission and absorption features ({\it yellow dotted lines}) is indicated, along with the masks ({\it grey bands}) superimposed to the emission of the UV airglow lines.}
    \label{fig:hst-spec}
\end{figure}

During this step, we also identified the presence of two strong emission features blueward of the Ly$\alpha$; the most intense one falls at the Ly$\alpha$ rest-frame wavelength of $1215.67$ \AA, whereas the position of the second one is compatible with that of the O {\footnotesize I} $+$ Si {\footnotesize II} system \citep{Vandenberk2001}. A visual inspection revealed that such features are single Gaussian-like lines rather than being formed by multiple components. 
These lines are present in all the three {\it HST} exposures, and are due to airglow emission originated as part of the UV sky background\footnote{See \url{https://hst-docs.stsci.edu/cosihb/chapter-7-exposure-time-calculator-etc/7-4-detector-and-sky-backgrounds}.}; since they are not relevant for the RBS\,1332 spectral analysis, we masked, thus excluded, them from any calculation. The RBS\,1332 average HST spectrum is shown in Fig.~\ref{fig:hst-spec}.

\begin{figure*}[htbp]
    \centering
    \begin{minipage}{0.49\textwidth}
        \includegraphics[width=\textwidth]{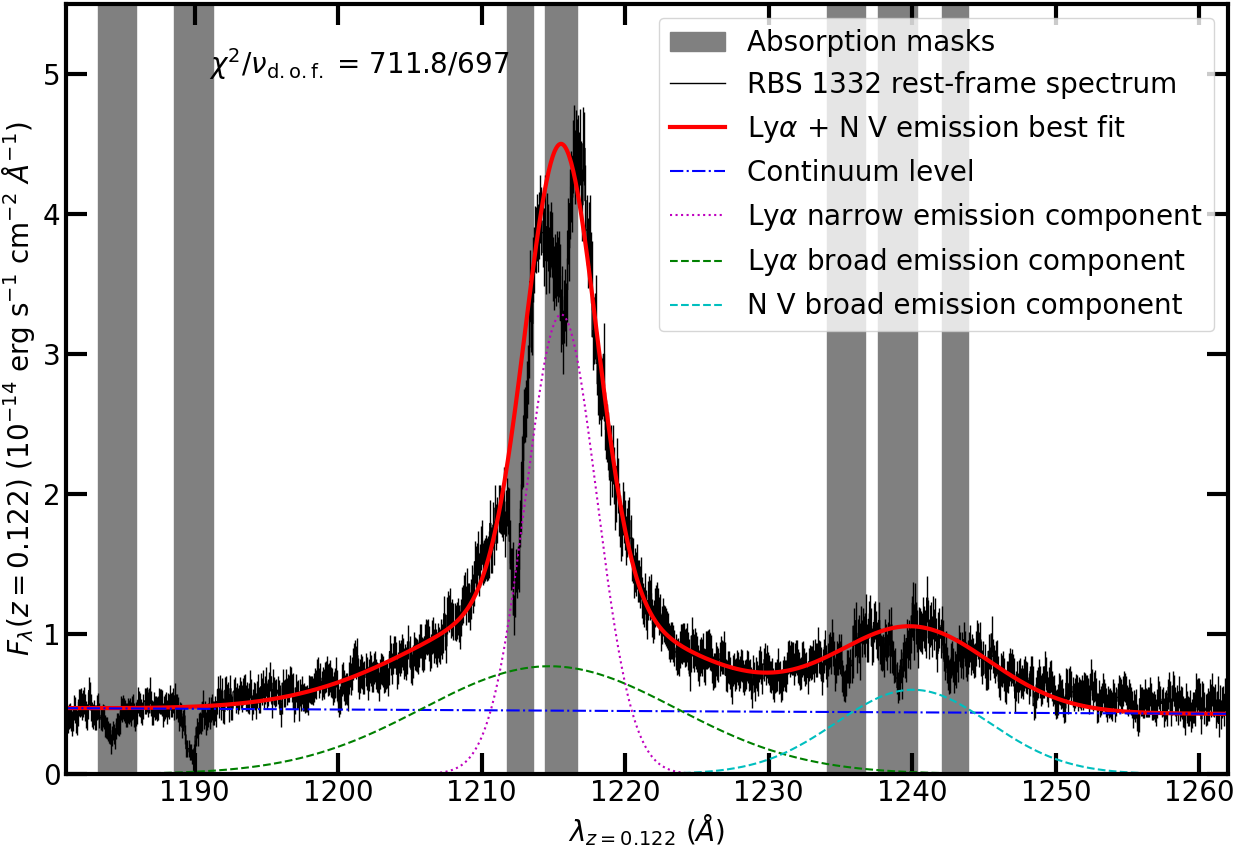}
    \end{minipage}
    \begin{minipage}{0.49\textwidth}
        \includegraphics[width=\textwidth]{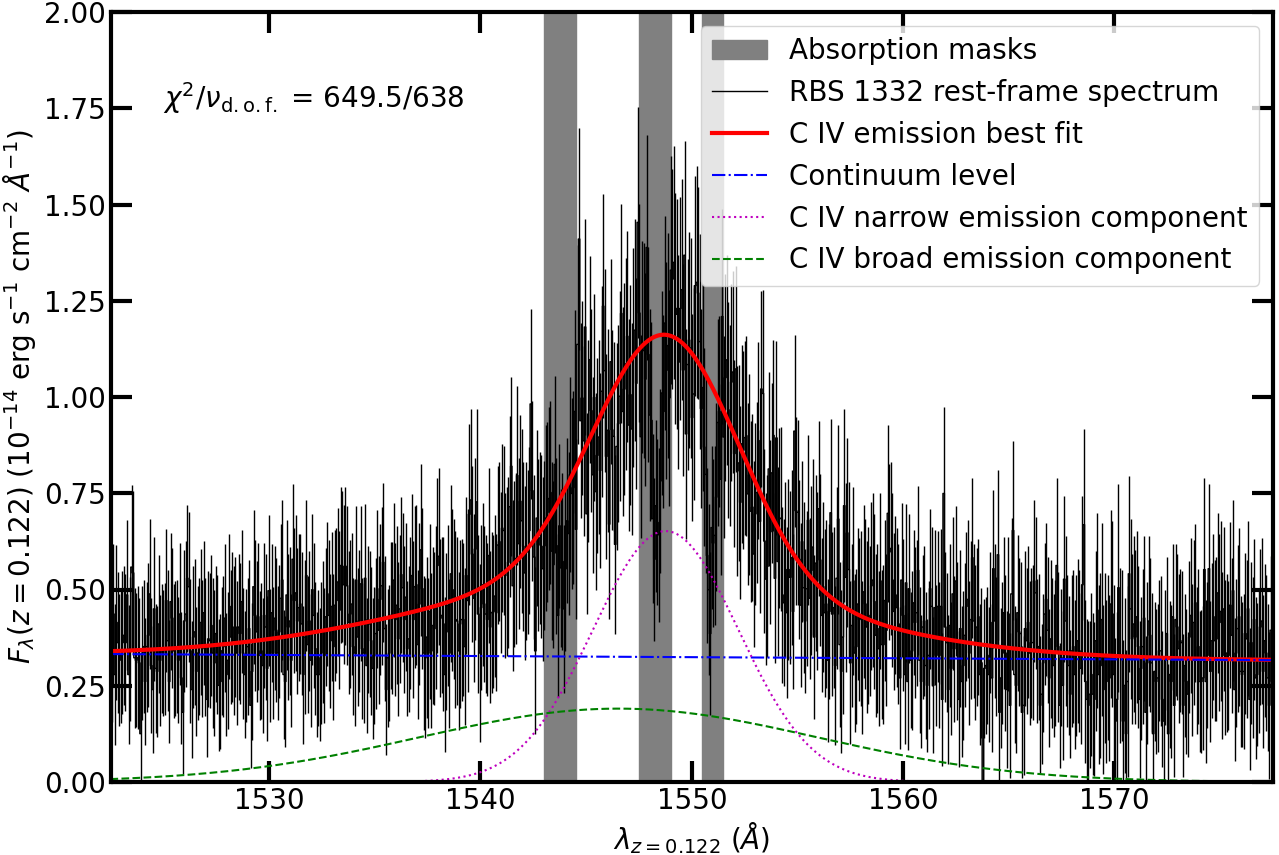}
    \end{minipage}
    \caption{{\it Left panel:} fit to the Ly$\alpha$+N {\scriptsize V} emission system of RBS\,1332 ({\it see insert}). {\it Right panel:} fit to the C {\scriptsize IV} emission ({\it see insert}). In both panels, the main absorption features identified through visual inspections ({\it grey bands}) are masked.}
    \label{fig:uv-emission}
\end{figure*}

We adopted the power-law shape found in Sect. \ref{sec:hst-ana} as our fiducial continuum over the entire spectral interval; on top of this component, we fitted the emission lines using multiple Gaussians:
\begin{itemize}
    \item a narrow and a broad component for both the Ly$\alpha$ and the C {\footnotesize IV};
    \item a single broad component for the N {\footnotesize V};
\end{itemize}
For all components, we left all the parameters free to vary; the goodness of each performed fit was evaluated by computing the $\chi^2$ value and the associated degrees of freedom number $\nu_{\rm d.o.f.}$. The result of such fits is shown in Fig. \ref{fig:uv-emission}, the obtained parameters of the line components are reported in Tab. \ref{tab:emi-pars}.

\begin{table}[h!t]
 \caption{Best-fit parameters of the RBS\,1332 Ly$\alpha$, N {\scriptsize V} and C {\scriptsize IV} emission lines, along with the associated 1$\sigma$ uncertainties.}
    \resizebox{\columnwidth}{!}{
    \centering
    \begin{tabular}{lcccc}
    \hline
    \multicolumn{5}{l}{ }\\
        Parameter & Ly$\alpha$ & N {\footnotesize V} & C {\footnotesize IV} & Units\\
    \multicolumn{5}{l}{ }\\
    \hline
    \multicolumn{5}{l}{ }\\
        $F_1$ & $4.27 \pm 0.14$ & $1.22 \pm 0.06$ & $1.20 \pm 0.24$ & $10^{-12}$ erg s$^{-1}$ cm$^{-2}$\\
        $F_2$ & $1.41 \pm 0.04$ & --- & $0.54 \pm 0.08$ & $10^{-12}$ erg s$^{-1}$ cm$^{-2}$\\
        $\Delta v_1$ & $-230 \pm 30$ & $-30 \pm 30$ & $-490 \pm 130$ & km s$^{-1}$\\
        $\Delta v_2$ & $-30 \pm 10$ & --- & $-60 \pm 30$ & km s$^{-1}$\\
        $\sigma_1$ & $2190 \pm 40$ & $1300 \pm 30$ & $1830 \pm 160$ & km s$^{-1}$\\
        $\sigma_2$ & $610 \pm 10$ & --- & $670 \pm 40$ & km s$^{-1}$\\
    \multicolumn{5}{l}{ }\\
    \hline
    \end{tabular}
    }
    \label{tab:emi-pars}
\end{table}

 It is immediate to note from Tab. \ref{tab:emi-pars} and Fig. \ref{fig:uv-emission} that we did not find evidence of narrow emission components ($\sigma < 600$ km s$^{-1}$) in any of the analyzed transitions. This feature is often found in objects that lie at the bright end of the AGN luminosity function \citep[see e.g.][and refs. therein]{Saturni2018}; in such cases, forbidden transitions -- such as the [O {\footnotesize III}] $\lambda$4959,5007 doublet -- tend to be weak, and an anti-correlation between their intensities and blended Fe {\footnotesize II} emission \citep{Boroson1992} is also expected. To investigate the possible weakness of the [O {\footnotesize III}] $\lambda$4959,5007 we retrieved the RBS\,1332 optical spectrum from the SDSS Data Release 16 \citep[DR16;][]{Blanton2017} portal\footnote{Available at \url{https://skyserver.sdss.org/dr16/en/tools/quicklook/summary.aspx}.}, and inspected the H$\beta$+[O {\footnotesize III}] spectral region: the [O {\footnotesize III}] is clearly detected, with both a line width typical of narrow-line regions (NLRs; $\sigma \sim 250$ km s$^{-1}$) and the evidence of a blue-shifted broader component, also indicative of the presence of an outflow similar to those discovered in other nearby AGN such as IRAS 20210+1121 \citep{Saturni2021}. Due to these findings, we conclude that the lack of narrow components in the high-ionization emission lines of RBS\,1332 must be ascribed to other physical phenomena, e.g. a lowly ionized or stratified NLR \citep[e.g.,][]{Wang2015}.

\subsection{AGN physical parameters of RBS\,1332}\label{sec:agn-phys}
From the decomposition of the RBS\,1332 UV spectrum, we derived all of the physical quantities that describe the AGN central engine. All the results of this paragraph are summarized in Tab. \ref{tab:agn-pars}, along with the corresponding 1$\sigma$ statistical uncertainties. We first computed the monochromatic luminosities at 1350 \AA\ $\lambda L_\lambda (1350\mbox{ \AA})$ and at 1450 \AA\ $\lambda L_\lambda (1450\mbox{ \AA})$ by integrating the RBS\,1332 spectrum in 100-\AA\ wide intervals centered around the rest-frame wavelength of interest; then, we used the C {\footnotesize IV} relation by \citet{Vestergaard2006} to derive the SMBH mass $M_{\rm BH}$ from $\lambda L_\lambda (1350\mbox{ \AA})$ and the FWHM of the narrower component of the C {\footnotesize IV} emission line:
\begin{equation}\label{eqn:mbh}
\resizebox{0.9\hsize}{!}{$
	\log{\left(
		\frac{M_{\rm BH}}{{\rm M}_\odot}
		\right)} = 6.66 + 2 \log{\left(
		\frac{{\rm FWHM}_{\rm CIV}}{1000\mbox{ }{\rm km}\mbox{ }{\rm s}^{-1}}
		\right)} + 0.53 \log{\left[
		\frac{\lambda L_\lambda (1350\mbox{ }{\rm \AA})}{10^{44}\mbox{ }{\rm erg}\mbox{ }{\rm s}^{-1}}
		\right]}
	$}
\end{equation}
In doing so, we note that our estimate of $M_{\rm BH}$ is fully compatible within the uncertainties with the SDSS DR16Q measurement of $\left(2.2^{+1.5}_{-0.9}\right) \times 10^7$ M$_\odot$ \citep[ see Tab. \ref{tab:agn-pars}]{Wu2022}; therefore, we decided to adopt it as our fiducial value without applying the correction by \citet{Denney2012} for C {\footnotesize IV} outflow-induced biases\footnote{Estimating $M_{\rm BH}$ with the \citet{Denney2012} relation from the C {\scriptsize IV} FWHM computed over the entire line profile would yield $(1.0 \pm 0.4) \times 10^7$ M$_\odot$, i.e. a factor of $\gtrsim$2 lower than the SDSS value.}. From the $M_{\rm BH}$ derived in this way, we estimated the Eddington luminosity $L_{\rm Edd}$ of RBS\,1332. Subsequently, we adopted the relation by \citet{Runnoe2012a,Runnoe2012b} to compute the AGN bolometric luminosity $L_{\rm bol}$ from $\lambda L_\lambda (1450\mbox{ \AA})$:
\begin{equation}\label{eqn:lbol}
    L_{\rm bol} = 0.75 \times 10^{4.745} \left[
    \lambda L_\lambda (1450\mbox{ \AA})
    \right]^{0.910}
\end{equation}
and thus the Eddington ratio $\varepsilon_{\rm Edd} = L_{\rm bol}/L_{\rm Edd}$.

Finally, we assumed the standard accretion disk model by \citet{Shakura1973} to estimate the maximum black-body temperature $T_{\rm max}$ of the disk assuming it reaches the innermost stable circular orbit:
\begin{equation}\label{eqn:tbb}
    T(r) = \left(
    \frac{3GM_{\rm BH}\dot{m}}{8\pi\sigma r^3}
    \right)^{1/4} \left(
    1 - \sqrt{\frac{R_{\rm ISCO}}{r}}
    \right)^{1/4}
\end{equation}
where $\dot{m}$ is the AGN accretion rate and $R_{\rm ISCO} = 6GM_{\rm BH}/c^2$ is the radius of the innermost stable orbit around the (non-rotating) central SMBH. To find $T_{\rm max}$, we estimated $\dot{m} = L_{\rm bol}/\eta c^2$ adopting a radiative efficiency $\eta = 0.057$ for a non-rotating SMBH \citep{Novikov1973}, and took the maximum value assumed by Eq. \ref{eqn:tbb} along the disk radius for our choice of free parameters. 

\begin{table}[h!t]
    \centering
    \caption{AGN parameters derived from the analysis of the RBS\,1332 C {\scriptsize IV} spectral region.}
    \label{tab:agn-pars}
    \begin{tabular}{lcr}
    \hline
    \hline
    \multicolumn{3}{l}{ }\\
    Quantity & Value & Units\\
    \multicolumn{3}{l}{ }\\
    \hline
    \multicolumn{3}{l}{ }\\
    ${\rm FWHM}_{\rm CIV}$ & $1570 \pm 90$ & km s$^{-1}$\\
    $\lambda L_\lambda (1350\mbox{ \AA})$ & $(2.1 \pm 0.6) \times 10^{44}$ & erg s$^{-1}$\\
    $\lambda L_\lambda (1450\mbox{ \AA})$ & $(2.0 \pm 0.6) \times 10^{44}$ & erg s$^{-1}$\\
    $L_{\rm bol}$ & $(0.8 \pm 0.2) \times 10^{45}$ & erg s$^{-1}$\\
    $L_{\rm Edd}$ & $(2.0 \pm 0.6) \times 10^{45}$ & erg s$^{-1}$\\
    $\varepsilon_{\rm Edd}$ & $0.4 \pm 0.2$ & ---\\
    $\dot{m}$ & $0.26 \pm 0.07$ & M$_\odot$ yr$^{-1}$\\
    $M_{\rm BH}$ & $(1.7 \pm 0.5) \times 10^7$ & M$_\odot$\\
    $R_{\rm ISCO}$ & $(1.5 \pm 0.4) \times 10^{13}$ & cm\\
    $T_{\rm max}$ & $(1.9 \pm 0.3) \times 10^5$ & K\\
    \multicolumn{3}{l}{ }\\
    \hline
    \end{tabular}
\end{table}

\subsection{Parameters of the absorption features associated with the Ly$\alpha$, N {\footnotesize V} and C {\footnotesize IV} transitions}\label{sec:agn-abs}
As a next step, we investigated the absorption features associated with the main UV transitions that are present in the RBS 1332 spectrum. To this aim, we first divided the spectrum by the pseudo-continuum obtained by summing all of the emission components derived in the previous steps; then, under the assumption that the AGN emission $F_\lambda^{(0)}$ is exponentially absorbed owing to the transport equation \citep[e.g.,][]{Rybicki1979}:
\begin{equation}\label{eqn:absflux}
    F_\lambda = F_\lambda^{(0)} \times e^{-\tau_\lambda}
\end{equation}
we empirically modeled the absorption optical depth $\tau_\lambda$ as a sum of Gaussian profiles:
\begin{equation}\label{eqn:optdep}
    \tau_\lambda = \sum_i A_i \exp{\left[
    -\frac{(\lambda - \lambda_i)^2}{2\sigma_i^2}
    \right]}
\end{equation}
The inspection of the Ly$\alpha$, N {\scriptsize V} and C {\footnotesize IV} spectral regions revealed that the Ly$\alpha$ absorption troughs are both formed by two narrower components; also the intermediate-velocity N {\scriptsize V} absorption is double, whereas the C {\footnotesize IV} troughs exhibit simpler single shapes. Therefore, we adopted four Gaussian profiles for both the Ly$\alpha$ and N {\scriptsize V} absorption, and only three for the C {\footnotesize IV} one, for a total of 11 components.

\begin{figure}[htbp]
    \centering
    \begin{minipage}{\columnwidth}
        \includegraphics[width=\textwidth]{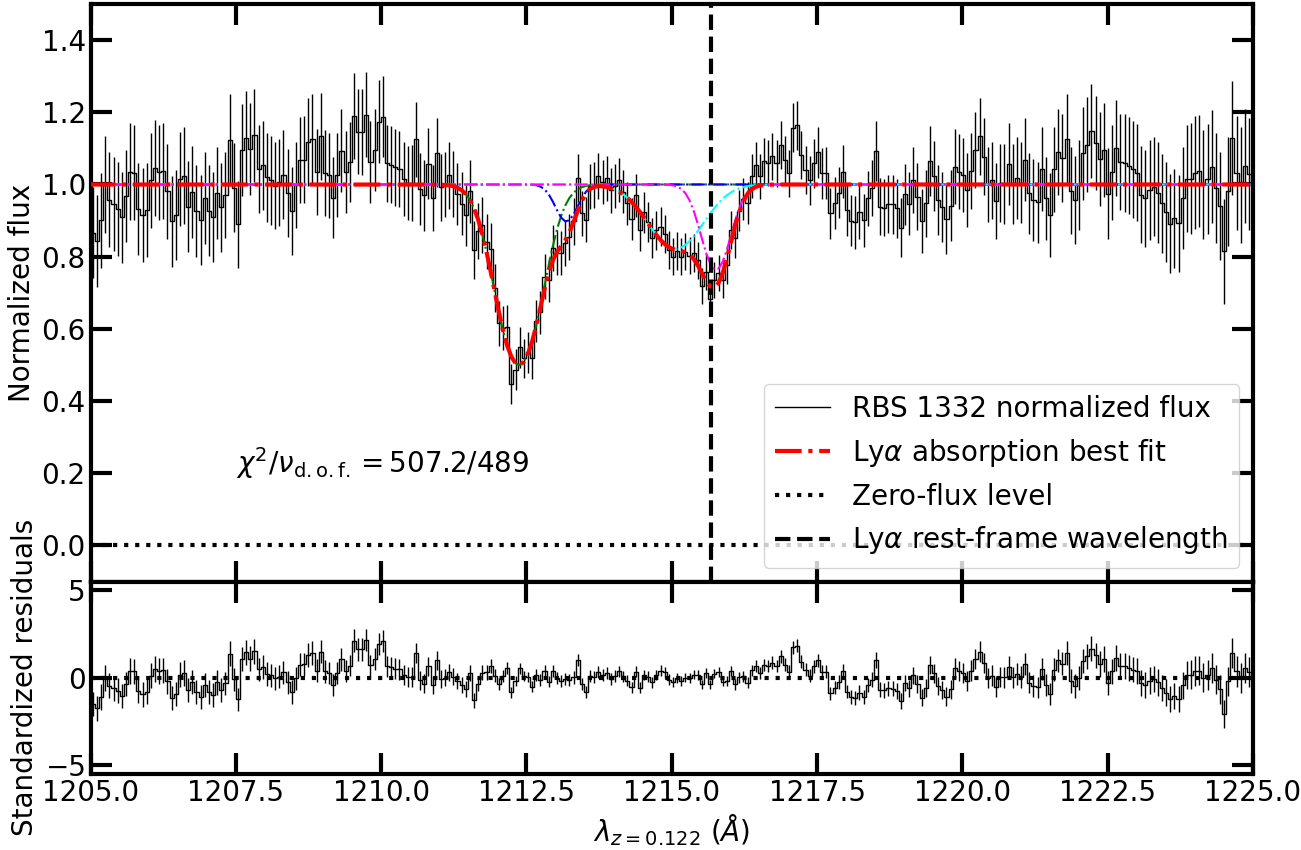}
    \end{minipage}
    \begin{minipage}{\columnwidth}
        \includegraphics[width=\textwidth]{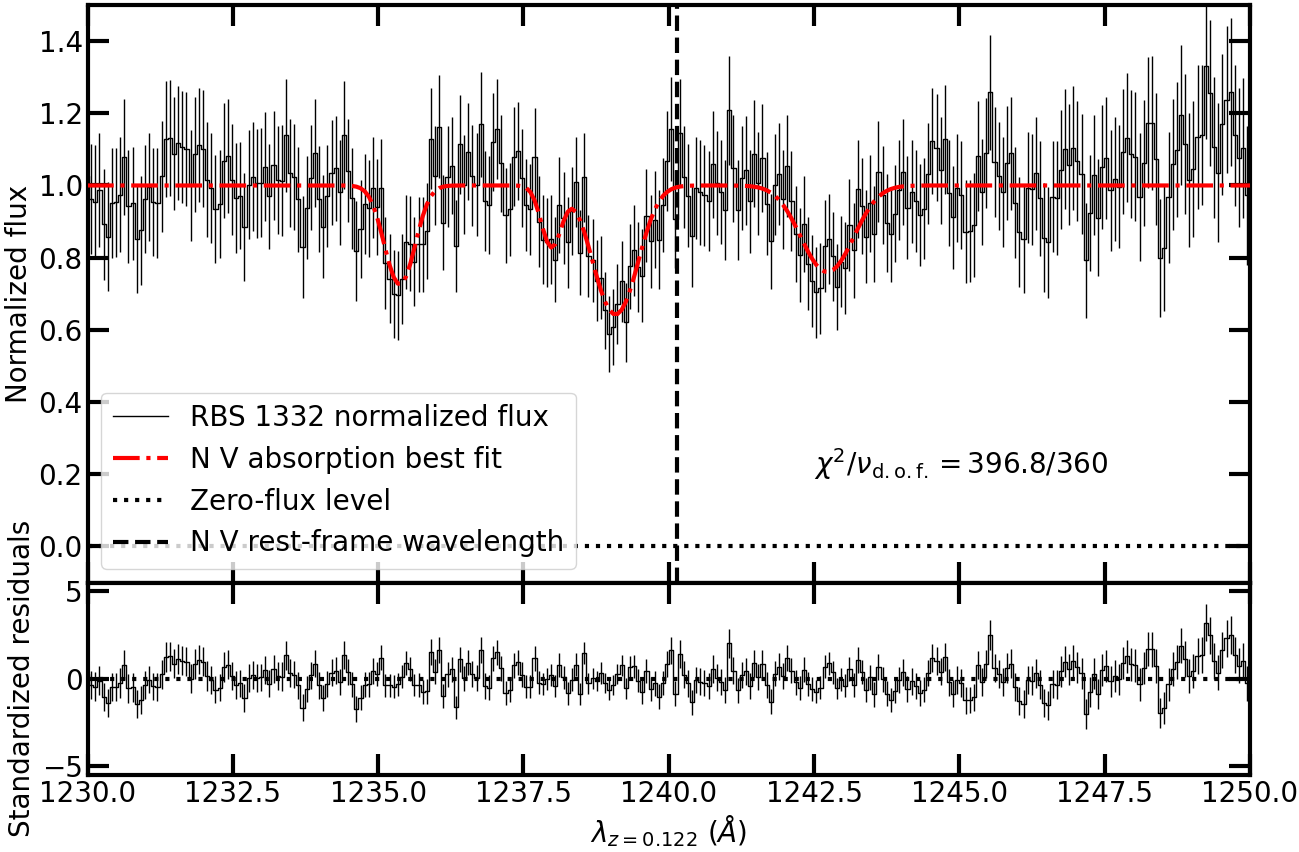}
    \end{minipage}
    \begin{minipage}{\columnwidth}
        \includegraphics[width=\textwidth]{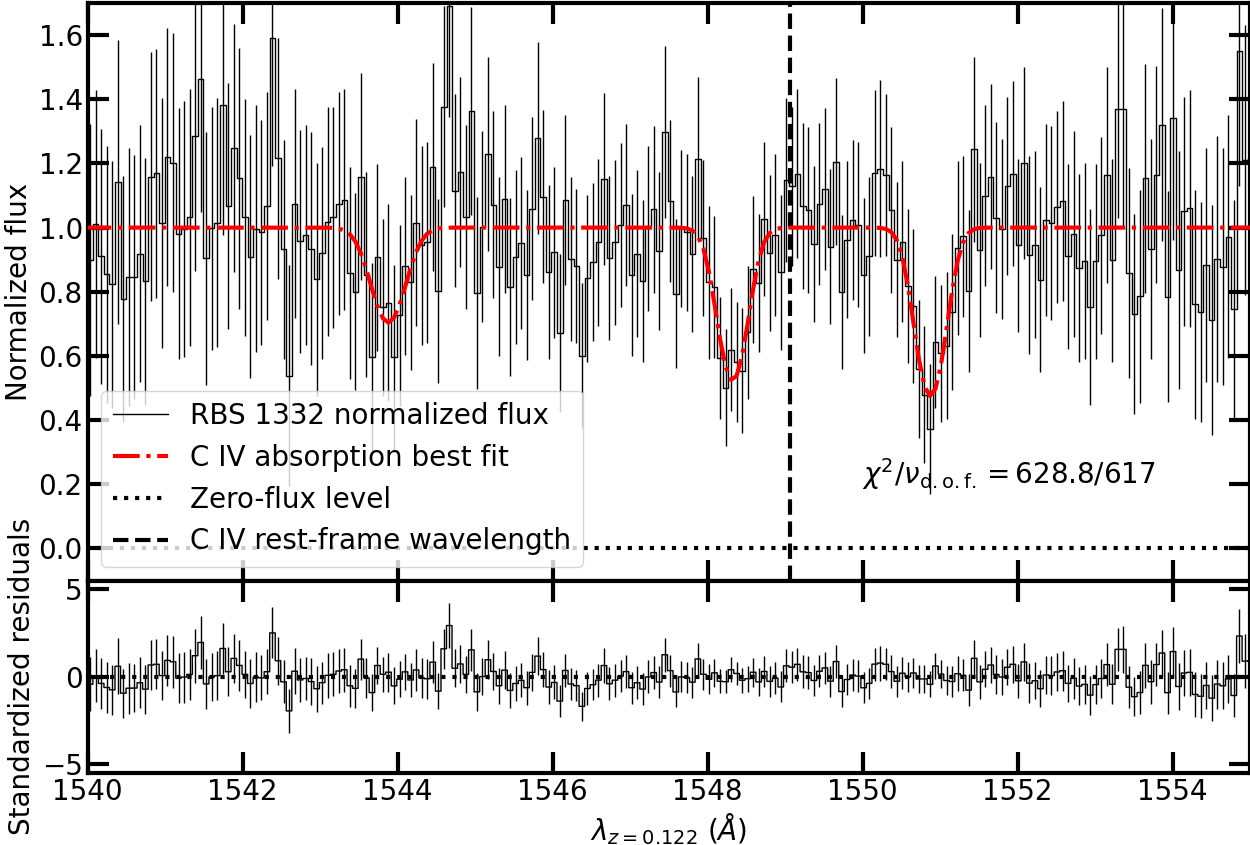}
    \end{minipage}
    \caption{Fits to the absorption systems ({\it red dot-dashed lines}) associated with the RBS\,1332 major emission features, along with the corresponding standardized residuals. {\it Upper panels:} fit to the Ly$\alpha$ absorption features. {\it Middle panels:} fit to the N {\scriptsize V} absorption features. {\it Lower panels:} fit to the C {\scriptsize IV} absorption features. In all panels, the zero-flux level ({\it dotted lines}) is indicated along with the rest-frame position of the relative emission feature ({\it dashed lines}).}
    \label{fig:abs-lines}
\end{figure}

We fitted this model to the normalized RBS 1332 spectrum, leaving all the absorption parameters free to vary. The best fits obtained in this way are shown in Fig. \ref{fig:abs-lines} for each spectral region, along with the corresponding $\chi^2$ and $\nu_{\rm d.o.f.}$ values. From the fit results, we computed the absorption equivalent widths ${\rm EW}_i$, velocity shifts $\Delta v_i$ and widths ${\rm FWHM}_i$ of each component, along with the corresponding 1$\sigma$ uncertainties; we report such values in Tab. \ref{tab:abs-pars}, numbering each absorption and grouping troughs associated to each transition according to their velocity shift (from blue to red). This led us to establish the following relations:
\begin{itemize}
    \item the most detached ``blue'' troughs (System 1) all have velocity shifts  in the range from $-1200$ km s$^{-1}$ to $-800$ km s$^{-1}$, with FWHMs in the range $300 \div 600$ km s$^{-1}$;
    \item the ``bluer'' intermediate troughs (System 2) have velocity shifts of $-500 \div -600$ km s$^{-1}$, with comparable FWHMs of $550 \div 650$ km s$^{-1}$ and also similar EWs (around $\sim$40 km s$^{-1}$ for both the Ly$\alpha$ and the N {\footnotesize V}); interestingly, the C {\footnotesize IV} transition does not show this absorption component;
    \item the ``redder'' intermediate troughs (System 3) exhibit similar velocity shifts of $-150 \div -250$ km s$^{-1}$, but more variegated FWHMs that range from $\sim$250 km s$^{-1}$ up to $\sim$1600 km s$^{-1}$;
    \item the ``red'' troughs (System 4) also have mixed properties in terms of both velocity shifts (from $\sim$0 to $\sim$600 km s$^{-1}$) and spread (FWHMs from $\sim$400 km s$^{-1}$ to $\sim$1200 km s$^{-1}$), being joined by mostly the fact of having positive velocity shifts with respect to the transition wavelength and by very similar EWs (all around $\sim$100 km s$^{-1}$).
\end{itemize}

Overall, the parameters reported in Table~\ref{tab:abs-pars} highlight how the analyzed absorption features exhibit intermediate velocity properties between broad \citep[BALs; e.g.,][]{Lynds1967,Weymann1991} and narrow absorption lines \citep[NALs; e.g.,][]{Reynolds1997,Elvis2000}, in line with those of the so-called mini-BALs \citep[see e.g.][and refs. therein]{Perna2024}. This interpretation is further supported by the lack of consistent variability over a maximum time interval $\Delta t \sim 13$ days between the UV observations, corresponding to $\sim$11.6 days in the rest frame; based on this evidence, under the assumption of a photoionization-driven absorption variability \citep{Barlow1992,Trevese2013}, we have been able to put an upper limit on the absorber's electron density $n_{\rm e}$ as:
\begin{equation}
    n_{\rm e} = \frac{1}{\alpha_{\rm rec} \Delta t_{\rm rf}}
\end{equation}
Assuming e.g. $\alpha_{\rm rec} = 2.8 \times 10^{-12}$ cm$^3$ s$^{-1}$ for the C {\footnotesize IV} troughs \citep{Arnaud1985} and $\Delta t_{\rm rf} > 11.6$ days, we got $n_{\rm e} < 3.6 \times 10^5$ cm$^{-3}$, consistent with the values typically expected for NAL systems \citep[e.g.,][]{Elvis2000,Saturni2016}.

To further quantify the RBS\,1332 absorption properties, we estimated the column densities $N_{\rm ion}$ associated with each trough \citep[e.g.,][]{Savage1991}. Under the assumption of an optically thin line lying on the linear part of the curve of growth associated to an absorber with full covering fraction, the minimum $N_{\rm ion}$ is related to its EW as \citep[e.g.,][]{Mehdipour2023}:
\begin{equation}\label{eqn:nion}
    N_{\rm ion}({\rm cm}^{-2}) \approx \frac{1.13 \times 10^{20}}{f \lambda^2} {\rm EW}({\rm \AA})
\end{equation}
where $f$ is the oscillator strength and $\lambda$ the laboratory wavelength of the transition. We retrieved the values of these quantities from the Atomic Spectra Database v5.12\footnote{Available at \url{https://www.nist.gov/pml/atomic-spectra-database}.} \citep{Kramida2024} provided by the National Institute of Standards and Technology (NIST). Based on the EW values that we found for the Ly$\alpha$, N {\footnotesize V} and C {\footnotesize IV} absorptions, we thus computed the corresponding minimum $N_{\rm ion}$ also reported in Tab. \ref{tab:abs-pars}. All $N_{\rm ion}$ derived in this way lie in the range 2--12 $\times 10^{13}$ cm$^{-2}$, in line with typical column densities found for objects with comparable absorption properties \citep[e.g.,][]{Wildy2016,Mehdipour2023}.

The similarities in the physical properties of these absorption systems hint at their common origin, potentially due to the existence of a clumpy \citep[e.g.,][]{Dannen2020,Ward2024} ionized outflow that is launched outward of the central engine by radiation pressure \citep[e.g.,][]{Murray1995,Proga2000,Risaliti2010}, then slows down and starts falling back into the AGN; alternatively, a scenario in which outflow velocities arise from the scattering-off material in a spiralling inflow \citep{Gaskell2016} may also work. Further multi-wavelength observations and detailed studies that are able to ({\it i}) derive the properties of the RBS\,1332 UV absorption from prime principles \citep[e.g., by computing numerical models with the {\footnotesize CLOUDY} software;][]{Ferland1998}, and ({\it ii}) relate them to other observable quantities such as the [O {\footnotesize III}] $\lambda$5007 broadening to infer the orientation of the central engine with respect to the line of sight \citep[e.g.,][]{Risaliti2011}, are needed to fully understand the physical processes at work behind such features.

\begin{table}[h!t]
    \centering
    \caption{{\it Upper section:} parameters of the RBS\,1332 absorptions associated to the Ly$\alpha$, N {\scriptsize V} and C {\scriptsize IV} transitions, along with the associated 1$\sigma$ uncertainties. In all columns, values are expressed in km s$^{-1}$. {\it Lower section:} minimum column density of the absorbing ions. In all columns, values are expressed in units of $10^{13}$ cm$^{-2}$.}
    \label{tab:abs-pars}
    \begin{tabular}{lccc}
    \hline
    \hline
    \multicolumn{4}{l}{ }\\
    Parameter & Ly$\alpha$ & N {\footnotesize V} & C {\footnotesize IV}\\
    \multicolumn{4}{l}{ }\\
    \hline
    \multicolumn{4}{l}{ }\\
    EW$_1$ & $98.2 \pm 3.0$ & $52.7 \pm 2.3$ & $79.9 \pm 3.6$\\
    EW$_2$ & $41.8 \pm 1.8$ & $40.4 \pm 1.8$ & ---\\
    EW$_3$ & $129.5 \pm 5.5$ & $72.1 \pm 2.9$ & $58.4 \pm 2.5$\\
    EW$_4$ & $94.3 \pm 4.2$ & $95.7 \pm 3.7$ & $105.4 \pm 4.0$\\
    $\Delta v_1$ & $-810 \pm 10$ & $-1160 \pm 10$ & $-1010 \pm 10$\\
    $\Delta v_2$ & $-600 \pm 10$ & $-530 \pm 10$ & ---\\
    $\Delta v_3$ & $-150 \pm 30$ & $-260 \pm 10$ & $-140 \pm 10$\\
    $\Delta v_4$ & $20 \pm 10$ & $630 \pm 10$ & $350 \pm 10$\\
    FWHM$_1$ & $270 \pm 10$ & $400 \pm 10$ & $610 \pm 10$\\
    FWHM$_2$ & $650 \pm 10$ & $540 \pm 10$ & ---\\
    FWHM$_3$ & $1620 \pm 40$ & $380 \pm 10$ & $250 \pm 10$\\
    FWHM$_4$ & $410 \pm 10$ & $1240 \pm 30$ & $1190 \pm 20$\\
    \multicolumn{4}{l}{ }\\
    \hline
    \multicolumn{4}{l}{ }\\
    $N_{\rm ion}^{(1)}$ & $5.48 \pm 0.17$ & $6.84 \pm 0.30$ & $6.81 \pm 0.31$\\
    $N_{\rm ion}^{(2)}$ & $2.33 \pm 0.10$ & $5.24 \pm 0.23$ & ---\\
    $N_{\rm ion}^{(3)}$ & $7.23 \pm 0.31$ & $9.35 \pm 0.38$ & $4.98 \pm 0.21$\\
    $N_{\rm ion}^{(4)}$ & $5.26 \pm 0.23$ & $12.42 \pm 0.48$ & $8.98 \pm 0.34$\\
    \multicolumn{4}{l}{ }\\
    \hline
    \end{tabular}
\end{table}

\section{X-ray Spectral analysis}

\begin{figure}
	\centering
	\includegraphics[width=\columnwidth]{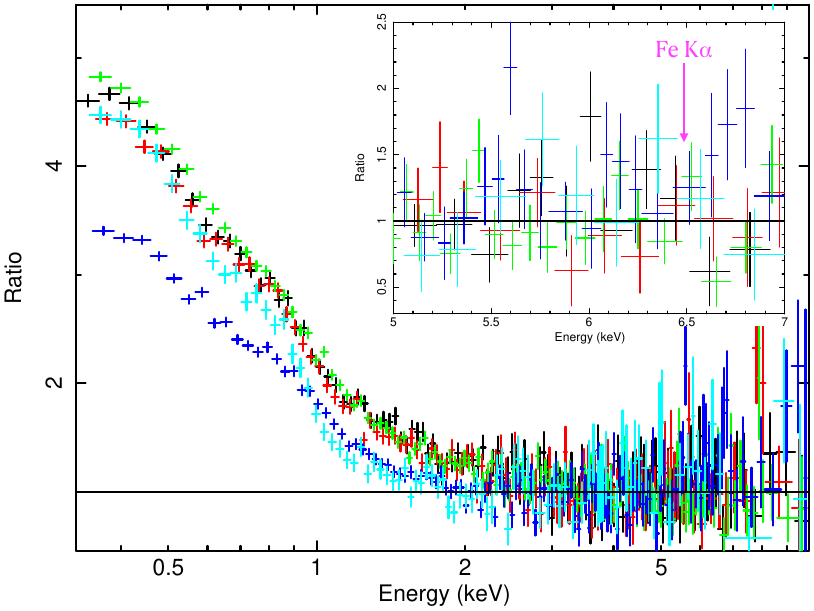}
	\caption{Ratios of the \textit{XMM-Newton} data to a power-law fitting the 3-10 keV energy range. A variable and remarkable soft-excess is clearly present below 3 keV, while no hints of a Fe K$\alpha$ emission line are observed.}
	\label{showsoft}
\end{figure}

In this section we will report on the broadband spectral fitting that were performed using \textsc{XSPEC} \citep[][]{Arnaud1996}. During the fitting procedure, the hydrogen column density due to the Milky Way $N_{\rm H}$=7.82$\times10^{19}$ cm$^{-2}$ \citep[][]{HI4PI2016} is always included and kept frozen to the quoted value. When COS and OM data are analysed, extinction is taken into account and we assumed E(B-V)=0.0067 \citep[][]{Schlafly2011}. We assumed the standard cosmology  $\Lambda$CDM framework with H$_{0}$=70 km s$^{-1}$ Mpc$^{-1}$, $\Omega_{\rm m}$=0.27 and $\Omega_{\lambda}$=0.73. 
\indent The X-ray emission of RBS\,1332 is dominated by a prominent and variable soft X-ray emission that exceeds the steep ($\Gamma>$2) primary continuum. In Fig.~\ref{showsoft} we show the residuals to a power-law component fitting only the 3-10 keV energy range. Interestingly, no features ascribable to a Fe K$\alpha$ emission line are observed, see inset in Fig.~\ref{showsoft}. Previous analysis aimed at characterizing the physical origin of the soft X-ray excess in AGN have focused on X-ray data only. However, such an approach has been found inconclusive as the main competing scenarios give similar best fit quality for the X-ray emission \citep[e.g.][]{Garcia2019}. However, they provide different estimates for the UV emission, thus, in the following, we will directly work on our broadband UV to X-ray dataset. \\

\begin{figure*}
	\centering
	\includegraphics[width=\columnwidth]{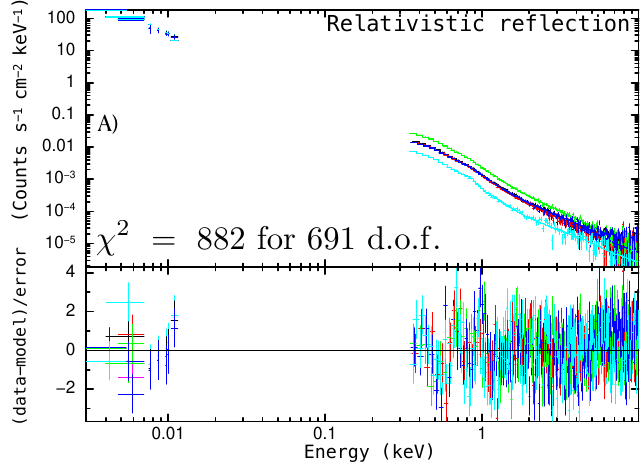}
	\includegraphics[width=\columnwidth]{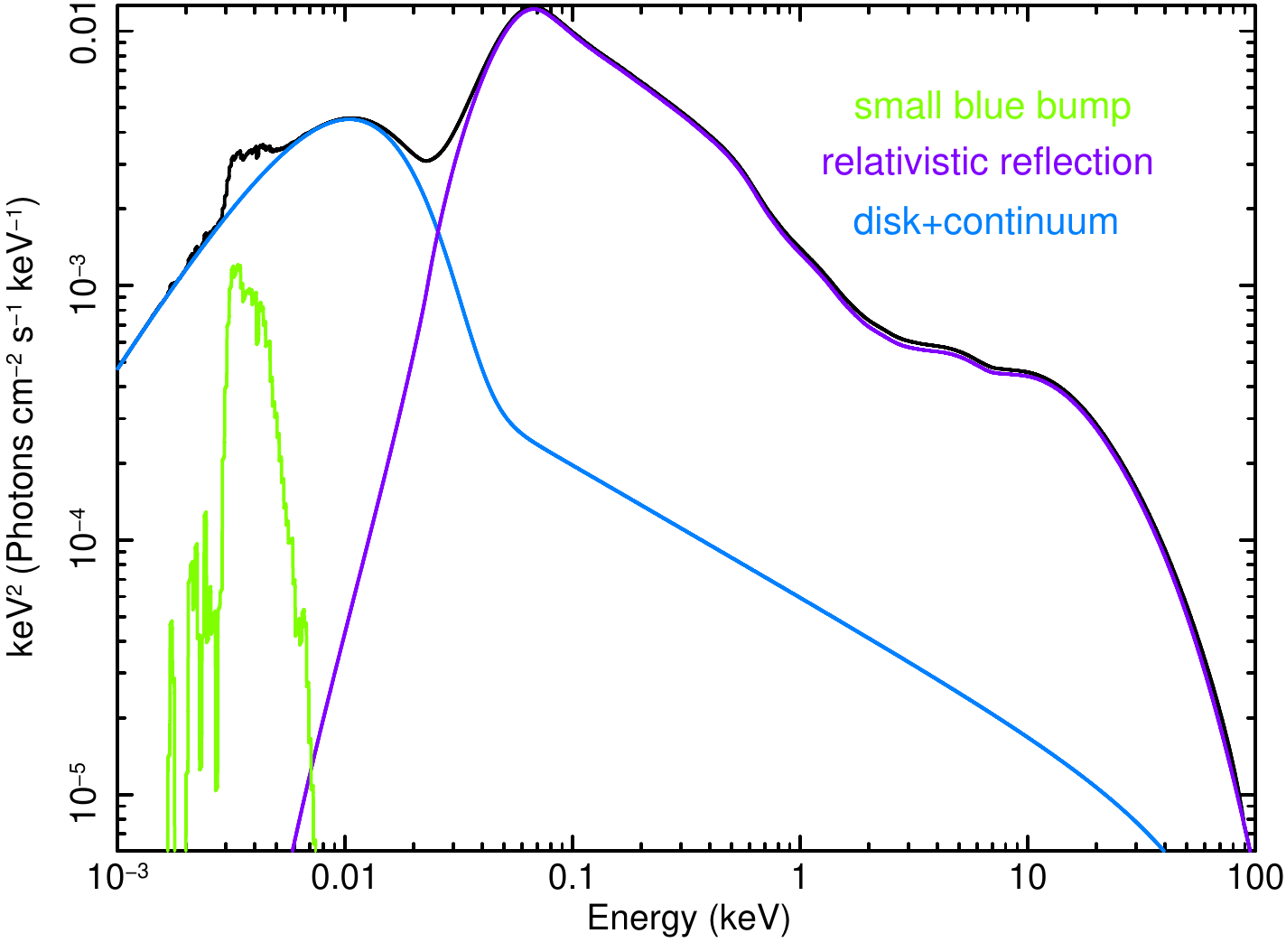}
    \includegraphics[width=\columnwidth]{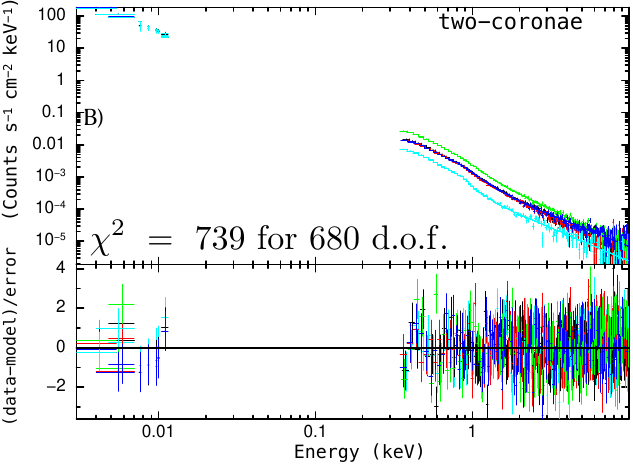}
	\includegraphics[width=\columnwidth]{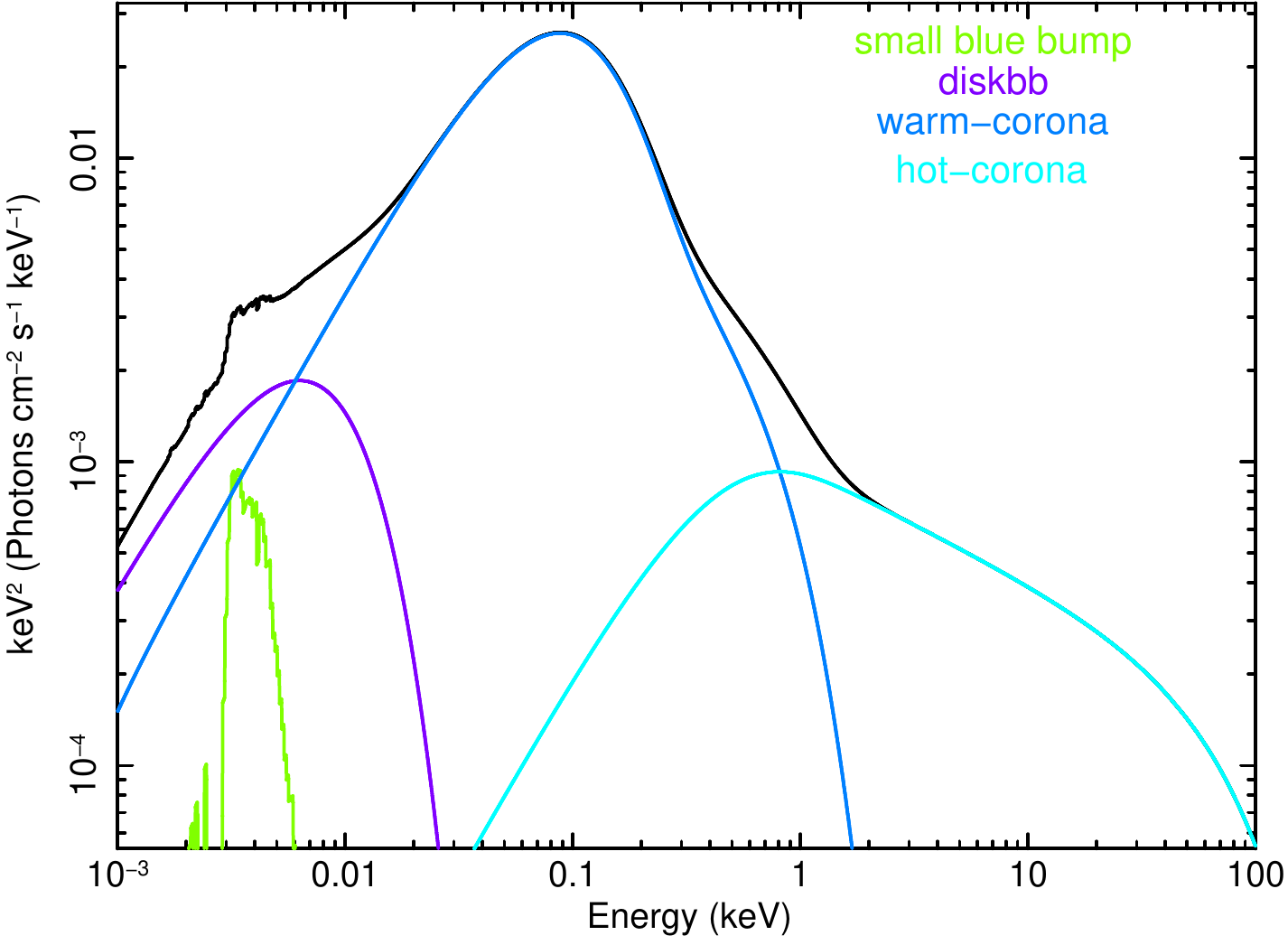}
	\caption{Top panels: Broadband fit to the OM, COS and EPIC-pn data assuming the soft X-ray excess to be dominated by relativistic reflection (top left ). On the top right panel we show the different contributions to the overall emission spectrum of RBS\,1332 for each of the model components as inferred from Obs. 1. Bottom panels: Broadband fit of the same dataset assuming the soft X-ray excess to be dominated by a warm Comptonization (left panel). Right panel: the different contribution to the overall emission spectrum of RBS\,1332 as derived from the first observation of our observational campaign.}
	\label{showrr}
\end{figure*}
\noindent \textbf{Testing relativistic reflection}:\\
\indent As stated, our dataset extends down to the UV domain, hence, in the modelling, we need to include any possible contribution of the Broad Line Region (BLR) which is responsible for excess of photons observed around 3000 $\AA$, the so-called small blue bump (SBB). We considered this spectral component using an additive table within \textsc{XSPEC} that was already presented in \citet[][]{Mehdipour2015}. During the fitting procedure, the normalization of the Small Blue Bump was free to vary (but tied among the observations). From the fit we obtained a flux of $F_{0.001-0.01~keV}=(8.7\pm0.3)\times10^{-13}$ erg s$^{-1}$ cm$^{-2}$ for this component. This flux is quantitatively compatible within the errors with the one obtaining by integrating the differential continuum derived in Sect. \ref{sec:hst-ana} from the {\it HST} data.\\
\indent Despite the lack of a significant Fe K$\alpha$ feature, we tested relativistic reflection as the origin of the X-ray soft excess in RBS\,1332. At this stage a reflection component cannot be ruled out as the lack of evidence of a Fe K feature can be explained by the presence of an extremely broad Iron line which is hard to detect in limited S/N data. Moreover, very weak Fe K emission line is expected when specific disk properties are met \citep[e.g. high ionization][]{Garcia2014}, or in presence of relativistic effects near a maximally spinning BH \citep[][]{Crummy2006} or simply by a very soft shape of the X-rays. \\
\indent Thus, we started modelling the hard X-ray continuum with \textsc{thcomp}, a convolution model meant to replace \textsc{nhtcomp} \citep[][]{Magdziarz1995,Zdziarski1996,Zycki1999}. This model describes a Comptonization spectrum by thermal electrons emitted by a spherical source and accounts for a sinusoidal-like spatial distribution of the seed photons, like in \textsc{compST} \citep[][]{Sunyaev1980}. \textsc{Thcomp} describes both upscattering and downscattering radiation and as a convolution model it can be used to Comptonize any seed photons distribution. In our case, we assumed the seed radiation to emerge from a \textsc{diskbb} component used to provide a standard spectrum from a multi-temperature accretion disk. Within \textsc{thcomp}, it is possible to compute the photon index $\Gamma$, the temperature of the seed photons and a covering fraction, $0\leq cov_{\rm frac} \leq 1$. When set to 1, all of the seed photons are Comptonized by the plasma while when this below 1, only a fraction of the seed photons is energized. In our computations we left $cov_{\rm frac}$ free to vary. We then linked to this continuum emission a relativistic reflection component using \textsc{relxillcp} \citep[e.g.][]{Garcia2014} and assumed the photon index of the two components to be the same. \textsc{Relxillcp}, combines the \textsc{xillver} reflection code and the \textsc{relline} ray tracing code \citep[e.g.][]{Dauser2016} and returns a spectrum due to irradiation of the accretion disk by an incident Comptonized continuum. Written in \textsc{XSPEC} notation, this model reads as:\\
\\
$\rm tbabs \times redden \times (SBB+thcomp \times diskbb+relxillcp)$.
\\
For each exposure of the monitoring, we calculated the source photon index $\Gamma$ (linking this parameter between \textsc{thcomp} and \textsc{relxillcp}), the temperature of the accretion disk (T$_{\rm bb}$ (eV)) and its normalization. For  \textsc{relxillcp}, the ionization parameter log $\xi$ (1/erg cm$^{-2}$ s$^{-1}$) and its normalization were fitted for each exposure. On the other hand, the iron abundance A$_{\rm Fe}$, the disc inclination {incl$^{\circ}$}, the BH spin (a) and the coronal emissivity (index1) were, instead, fitted tying their values among the observations. The covering fraction of the Comptonizing plasma was also fitted linking its value among the pointings.\\
 \indent This approach led us to a fit statistic $\chi^2$=882 for 691 d.o.f. and we show the corresponding fit and its accompanying quantities in Fig.~\ref{showrr} and Table~\ref{tablerr}, respectively.

\begin{table*}
	\centering
	\caption{Best-fit parameters derived for the relativistic reflection scenario. Errors are given at the 90\% confidence level for the single parameter of interest.} \label{tablerr}
	\label{tabtext}
	\setlength{\tabcolsep}{4.5pt}
	\begin{tabular}{l l c c c c c }
\hline
\hline \noalign{\smallskip}
Relativistic reflection&&&&&& \\
&&Obs. 1 & Obs. 2 &Obs. 3 &Obs. 4  &Obs. 5  \\
SBB&Norm$\dagger$ ($\times10^{-3}$)& 2.7$\pm$0.1& & &  & \\
redden&E(B-V)$\dagger$&0.0067& & &  & \\
Tbabs&N$_{\rm H}\dagger$ ($\times10^{19}$ cm$^{-2}$) & 7.82& & &  & \\
thcomp&$\Gamma$& 2.52$\pm$0.02&2.52$\pm$0.03&2.55$\pm$0.01&2.41$\pm$0.02&2.51$\pm$0.03\\
&CF$\dagger$& 7\%$\pm$4\%& & & &\\
Diskbb&T$_{\rm bb}$ (eV)& 4.7$\pm$0.9&4.8$\pm$0.8 &3.7$\pm$0.6 &4.6$\pm$0.6 &5.0$\pm$0.7 \\
&Norm ($\times10^{9}$)& 1.4$\pm$0.6&1.3$\pm$0.8 &2.9$\pm$1.1 &1.5$\pm$0.7 &1.1$\pm$0.7 \\
relxillcp&incl& (58$\pm$2)$^\circ$& & & & \\
&index1& >8.85& & & & \\
&a& >0.997& & & & \\
&A$_{\rm Fe}$& 1.6$\pm$0.3& & & &\\
&$\log \xi$&3.12$\pm$0.05&3.10$\pm$0.05 &3.20$\pm$0.05 &3.00$\pm$0.06& 2.7$\pm$0.1 \\
&Norm ($\times10^{-5}$)& 2.6$\pm$0.1 &2.4$\pm$0.1&4.5$\pm$0.2 &2.4$\pm$0.1 &1.7$\pm$0.2 \\
\hline
\end{tabular}
\end{table*}

In accordance with this model, the broadband emission of RBS\,1332 can be ascribed to a black-body like emission that dominates the FUV and UV wavelengths. No significant changes in its temperature are observed. In this depicted scenario, only a marginal fraction of the photons emitted by the disk are Comptonized by the hot corona (${\rm cov}_{frac}\sim7\%$) and the X-rays can be mainly ascribed to extreme relativistic reflection. A very steep emissivity is found (index1 >8.85), this implying the Comptonizing plasma to be very close to the maximally rotating SMBH (a>0.997).

\noindent \textbf{Testing warm Comptonization:}\\
\indent Alternatively to blurred ionized reflection we tested the two-coronae model \citep[e.g.][]{Petrucci2018,Kubota2018}. Within this framework, the broadband emission spectrum of AGN is accounted for by distinct emitting zones, a standard accretion disk; a warm Comptonizing plasma; and the inner hot corona. In particular the disk can be either non-dissipative and in this case be fully covered by the warm corona or, instead, contribute to the overall flux of the source. In this second scenario the warm corona is assumed to be patchy and part of the energy dissipated in the disk leaked through the warm coronal region. The warm corona is defined as an optically thick geometrically thin medium in which Comptonization is the dominant cooling mechanism \citep[][]{Petrucci2018,Petrucci2020} while the hot corona is the one standard depicted in the two-phase model. We thus modified our previously adopted test model as follows:\\

\noindent $ \rm tbabs \times reddend \times (SBB+thcomp_{W} \times diskbb+ nthcomp_{H}$).\\

Thus we model the broadband emission spectrum of RBS\,1332 assuming an outer disk for which we compute its temperature (T$_{bb}$) linking its value among the observations. The normalization Norm$_{\rm disk}$ is instead derived in each observation. Then, we assume the soft X-rays emerging from a closer region in which a fraction of disk photons are Comptonized ($\rm thcomp_W \times diskbb$). For this second component we fitted the photon index, the temperature and normalization of the disk photons, the covering fraction of the warm corona and its normalization. Finally, for the hot corona we assumed that the seed photons cooling it are only those from the warm component (T$_{bb}^{H}=kT_{\rm warm}$). We fixed the hot corona temperature to a constant value of kT= 50 keV, in agreement with average estimates \citep[e.g.][]{Middei2019b,Kamraj2022}. Thus, we only fitted the photon index $\Gamma_{H}$ and the model normalization. These simple steps led us to a best-fit of $\chi^2$/d.o.f.=739/680 shown in Fig.~\ref{showrr} bottom panels.

\begin{table*}
	\centering
	\caption{Best-fit parameters corresponding to the two-coronae fitting model. Fluxes and luminosity are the observed ones.} \label{tabella}
	\label{tabtext}
	\setlength{\tabcolsep}{4.5pt}
	\begin{tabular}{l l c c c c c }
		\hline
		\hline \noalign{\smallskip}
		warm Comptonization&&&&&& \\
		&&Obs. 1 & Obs. 2 &Obs. 3 &Obs. 4  &Obs. 5  \\
  SBB&Norm$\dagger$ ($\times10^{-3}$)& 2.7$\pm$0.1& & &  & \\
redden&E(B-V)$\dagger$&0.0067& & &  & \\
Tbabs&N$_{\rm H}\dagger$ ($\times10^{19}$ cm$^{-2}$) & 7.82& & &  & \\
	diskbb & T$_{in}\dagger (eV)$ & 1.2$\pm$0.2 & & & & \\
       & Norm ($\times10^{11}$)& 1.0$\pm$0.4 & 1.0$\pm$0.3 & 1.2$\pm$0.3&1.1$\pm$0.4 &0.9$\pm$0.3\\
warm corona&  $\Gamma_{\rm soft}$ &2.65$\pm$0.20 &2.52$\pm$0.19&2.41$\pm$0.15 &2.59$\pm$0.09&2.48$\pm$ 0.16\\
&kT$_{\rm e}$ (keV)& 0.19$\pm$0.03&0.19$\pm$0.02 &0.18$\pm$0.01&0.24$\pm$0.02&0.17$\pm$0.01 \\
        &$\tau_{\rm wc}$ & 30$\pm$3&32$\pm$6 &35$\pm$4&28$\pm$3&35$\pm$3 \\
       & cov$_{\rm frac}$ &0.18$\pm$0.07 & 0.15$\pm$0.06 & 0.19$\pm$0.05 &0.17$\pm$0.04 &0.19$\pm$0.08 \\
       & T$_{\rm bb} (eV)$ &35$\pm$2 & 33$\pm$3 & 39$\pm$2 & 33$\pm$2 &23$\pm$7 \\
       & Norm ($\times10^{6}$)& 3.2$\pm$0.8 &3.7$\pm$1.2 &2.2$\pm$0.7&3.8$\pm$0.8 &11$\pm$5 \\
 hot corona& $\Gamma_{\rm hot}$ &2.39$\pm$0.09 & 2.34$\pm$0.09&2.36$\pm$0.03&2.01$\pm$0.09&2.1$\pm$0.1 \\
      & Norm ($\times10^{-4}$)&9.5$\pm$1.0 &8.51$\pm$0.8&15.1$\pm$0.9&6.6$\pm$0.8&3.0$\pm$0.3 \\
      \hline 
    &L$_{\rm 2-10~keV}$ ($\times10^{43}$ erg s$^{-1}$) &5.6$\pm$0.2&5.3$\pm$0.2&9.1$\pm$0.2&6.7$\pm$0.3&2.5$\pm$0.3\\
     &   F$_{\rm 0.5-2~keV}$ ($\times10^{-12}$ erg cm$^{-2}$ s$^{-1}$)&3.5 $\pm$0.3&3.2$\pm$0.3&5.9$\pm$0.4&3.5$\pm$0.3& 1.5$\pm$0.6 \\
      &      F$_{\rm 2-10~keV}$ ($\times10^{-12}$ erg cm$^{-2}$ s$^{-1}$)&1.4$\pm$0.2&1.3$\pm$0.1&2.3$\pm$0.2&1.7$\pm$0.1&0.7$\pm$0.1 \\ \hline 
      	\end{tabular}
\end{table*}

\begin{figure}
	\centering
	\includegraphics[width=\columnwidth]{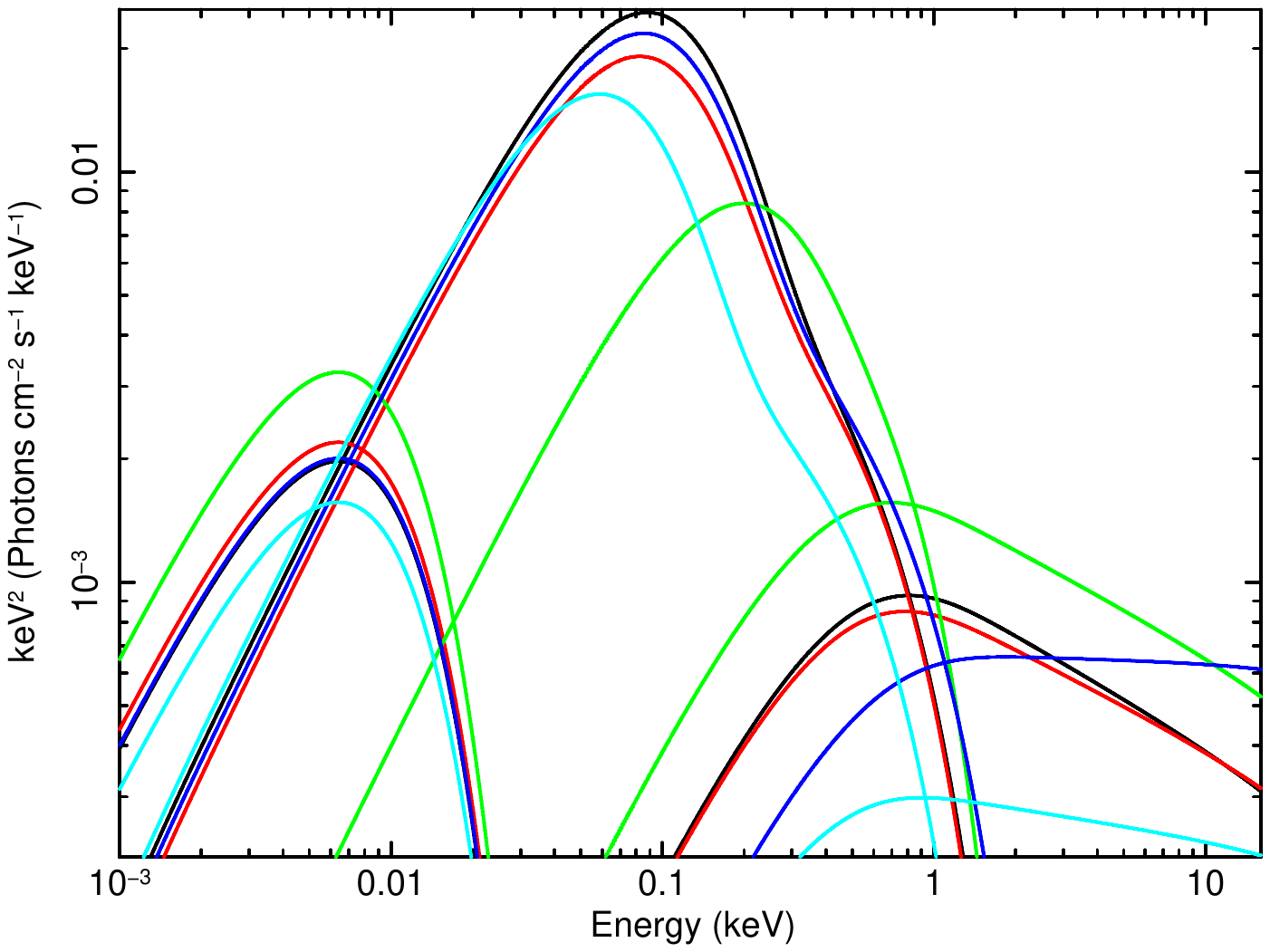}
	\caption{The three components shaping the broadband emission of RBS\,1332 are shown for each observation of the campaign.}
	\label{plotcomp}
\end{figure}

This best-fit to the data agrees with the optical-to-X-rays emission of RBS\,1332 emerging from three distinct components. A hot corona with a  soft spectral index ($\Gamma_{\rm hard}\sim$2.2). Then a warm corona region characterized by an average spectral index steeper than the primary continuum ($\Gamma_{\rm soft}\sim$2.65) and an average temperature kT$_{\rm warm}\sim$0.2 keV. These values can be translated into a Thompson opacity\footnote{This computation relies on the equation $\tau=(2.25+3/(\theta\times((\Gamma+0.5)^2-2.25)))^{1/2}-1.5$  in which a spherical geometry of the Comptonizing medium is assumed. This comes from an internal routine within the \textsc{nthcomp} model.} of $\tau_{\rm warm}\sim$30. These inferred parameters for the warm corona are in full agreement with the bulk of measurements reported by \citet{Petrucci2018}. Finally, we add a standard disk responsible for the optical emission, part of which is Comptonized by the hot corona.
In Tab. ~\ref{components} we show the relative flux of the various components used to model the UV-to-X-ray spectra of RBS\,1332 and in Fig.~\ref{plotcomp} we show the corresponding emission components. The best-fit temperature of the accretion disk $T_{\rm in} \sim 1.2$ eV is a factor of $\sim$10 less than the $T_{\rm max}$ reported in Tab. \ref{tab:agn-pars}; however, we remark that $T_{\rm max}$ is just the upper limit to the range of BB temperatures achievable by an optically thick and geometrically thin accretion disk \citep{Shakura1973}. Furthermore, the $T_{\rm max}$ derived from the UV spectral analysis is only a factor of $\sim$2 lower than the average $T_{\rm bb} \sim 30$ eV -- corresponding to $\sim$3.5 $\times 10^5$ K -- found for the BB component of the warm corona. Based on these evidences, we conclude that all the gas temperatures obtained from the UV and X-ray analyses are in line with the typical values expected for accretion-powered AGN activity.

\begin{table}
	\centering
	\caption{Fluxes for the different emission components adopted to model the broadband emission of RBS\,1332 assuming the two-coronae model.} \label{components}
	\label{tabtext}
	\setlength{\tabcolsep}{4.5pt}
	\begin{tabular}{l c c c }
		\hline
		\hline \noalign{\smallskip}
& diskbb &warm corona &hot corona \\
&(0.001-0.3~keV) &0.3-2~keV &2-10~keV\\
\hline
Obs. 1 &-11.20$\pm$0.02 &-11.20$^{+0.12}_{-0.02}$ &-11.91$\pm$0.09\\
Obs. 2 &-11.19$\pm$0.02 &-11.40$\pm$0.05 &-11.87$\pm$0.02\\
Obs. 3 &-11.19$\pm$0.02 &-11.10$\pm$0.02 &-11.63$\pm$0.01\\
Obs. 4 &-11.19$\pm$0.02&-11.30$\pm$0.05 &-11.75$\pm$0.02\\
Obs. 5 &-11.19$\pm$0.02 &-11.67$\pm$0.03  &-12.18$\pm$0.03  \\
       \end{tabular}
\end{table}
In the context of the two-coronae model we find that the disc component is fairly constant during the campaign while both the hard power-law and the warm corona changes in flux by a compatible amount. As a final test, we fit the broadband data of RBS\,1332 using the model \textsc{AGNSED}, \citep[][]{Done2012,Kubota2018}. Within this model, the three distinct emitting zones (outer disk, warm corona and hot corona) are energetically coupled and assumed to be radially stratified. The first one emits as a standard disc black body (BB) from R$_{\rm out}$ to R$_{\rm warm}$, as warm Comptonization from R$_{\rm warm}$ to R$_{\rm hot}$ \citep[see][for details on this region]{Petrucci2018,Petrucci2020} and, below R$_{\rm hot}$ down to R$_{\rm ISCO}$ as the typical geometrically thick, optically thin electron distributions expected to provide the AGN hard X-ray continuum \citep[][]{Haardt1993}. Radii are in units of gravitational radii $R_{\rm g}$. In the fitting procedure, we calculated the hard and the warm corona photon indices ($\Gamma_{\rm hard}$ and $\Gamma_{\rm soft}$), the warm coronal temperature $kT_{\rm e}$ (keV) while the hard coronal temperature was kept to a fixed value of 50 keV. A co-moving distance of 539 Mpc was derived from the redshift and fixed in the model. We assumed a SMBH mass of $\log$ M$_{BH}$=7.2$\pm$0.2, based on our computations from the UV spectrum in Sect. 3. We set the scale height for the hot Comptonization component, HT$_{\rm max}$ to 10 gravitational radii R$_g$, this mimicking a spherical Comptonizing plasma of similar radius. Finally, we assumed a disk inclination of 30$^\circ$ and a spin of 0.5. This led us to a best-fit of $\chi^2$=785 for 691 d.o.f., fully compatible with the one previously obtained. The spectral index of the hot and warm Comptonization regions are in good agreement with those in Table~\ref{tabella} found in the previous warm Comptonization fit. This fit provides estimates of the size of the different regions of emission. The hot corona appears quite compact ($\sim10~R_{\rm g}$), while the warm Comptonization area extends up to $\sim 200~R_{\rm g}$. We note however variations of this radius $R_{\rm warm}$ of a factor 4 in the 5 observations while the hot coronal region remains more constant. The derived Eddington ratio is in the range is between $\sim25-55\%$ Eddington.\\

\begin{table}
	\centering
	\caption{Radial extension in gravitational radii derived from \textsc{AGNSED}.} 
	\label{tabAGNSED}
	\setlength{\tabcolsep}{0.5pt}
	\begin{tabular}{l l c c c c c }
		\hline
		\hline \noalign{\smallskip}
		&&Obs. 1 & Obs. 2 &Obs. 3 &Obs. 4  &Obs. 5  \\
  		&R$_{\rm hot}$ & 12$\pm$2&10$\pm$2& 12.5$\pm$0.5&9$\pm$1  &7$\pm$1\\
  		&R$_{\rm warm}$ & 200$\pm$35&190$\pm$80& 320$\pm$40&175$\pm$30  &80$\pm$20\\
     	&$\log \dot{m}$ & -0.41$\pm$0.03&-0.42$\pm$0.07&-0.27$\pm$0.01&-0.42$\pm$0.03&-0.58$\pm$0.04 \\
	\end{tabular}
\end{table}

\noindent \textbf{Warm corona or relativistic reflection?}\\
\indent We complement the analyses presented in previous Sect.s 4.1 and 4.2 testing \textsc{reXcor}\footnote{\textsc{ReXcor} is freely available for spectral fitting using \textsc{XSPEC} at the webpage \url{https://github.com/HEASARC/xspec_localmodels/tree/master/reXcor}.} \citep[][]{Xiang2022,Ballantyne2024}, a spectral model that self-consistently calculates the effects of both ionized relativistic reflection (incorporating the light bending and line's blurring) and the emission from a warm corona. Within \textsc{reXcor}, the accretion power released in the inner regions of the accretion disk (<400 $\rm r_g$) is distributed among a hot corona (for wich a lamppost geometry is assumed), a warm corona and the accretion disk itself. The model \textsc{reXcor} accounts for a plethora of shapes for the soft X-ray excess that depends on the amount of energy dissipated either by the lamppost or the warm corona.\\

\indent This model can be used in \textsc{XSPEC} as additive tables that were computed for fixed values of BH spin (0.9,0.99), lamppost height (5 R$_{\rm g}$,20 R$_{\rm g}$) and Eddington ratio (1\%,10\%). Up to 5 parameters can be fitted in these tables: the fraction of the accretion flux dissipated in the lamppost ($f_{\rm x}$), the photon index of the power-law continuum $(\Gamma)$, the fraction of the accretion flux dissipated in the warm corona ($h_{\rm f}$); the warm corona opacity $\tau$, and the model normalization.
Then, we tested on our dataset the following \textsc{XSPEC} model:\\

$\rm tbabs \times reddend \times (diskbb+nthcomp+reXcor(\lambda=0.1, a=0.99, h=5 r_{g}) )$. \\

We fitted the temperature of the \textsc{diskbb} component linking its value among the observations while its normalization was  computed for each dataset. The fraction of the energy dissipated in the lamppost, the warm corona as well as the opacity of this second component and the model normalization were free to vary and fitted for each epoch.
We tied the photon index of \textsc{reXcor} to the one of  \textsc{nthcomp} and computed it. Concerning \textsc{nthcomp}, we assumed its T$_{(\rm bb)}$ temperature to be the same as the disk while the temperature of  the hot electron were fixed to 50 keV. 
We started fitting the \textsc{reXcor} table considering the full \textit{XMM-Newton} band, but due to the limited gamma range ( 1.5$<\Gamma<$2.2) allowed in the model, it struggles to account for the very first soft bins of the spectra. Consequently, we excluded data below 0.5 keV and refitted the spectra, obtaining  a better fit statistic $\chi^2=$760/663, which is statistically compatible with the one obtained with the two-coronae model. We found that our choice of ignoring data below 0.5 keV does not modify the best fit values for the fraction of energies dissipated in the lamppost or the warm corona.
For the sake of simplicity, in Table~\ref{tabsreXcor} we only report the quantities derived for the \textsc{reXcor} table, as the other parameters are consistent within the errors with the previously obtained values.

\begin{table}
	\centering
	\caption{Results of the \textsc{reXcor} model on the RBS\,1332 dataset.} 
	\label{tabsreXcor}
	\setlength{\tabcolsep}{1.5pt}
	\begin{tabular}{l l c c c c c }
		\hline
		\hline \noalign{\smallskip}
		&&Obs. 1 & Obs. 2 &Obs. 3 &Obs. 4  &Obs. 5  \\
          &$\Gamma$& >2.19&2.18$\pm$0.01&2.18$\pm$0.01 &>2.19&>2.19\\
        &$f{\rm x}$& 0.02$\pm$0.01&<0.06&<0.02&0.05$\pm$0.01&>0.14 \\
  		&h$_{\rm f}$ & 0.76$\pm$0.01&0.65$\pm$0.06&0.76$\pm$0.01&0.71$\pm$0.02&0.55$\pm$0.05\\
  		&$\tau$ & >29&21$\pm$4&23$\pm$3&>28.6&27$\pm$2\\
	\end{tabular}
\end{table}

The use of the \textsc{reXcor} model supports the interpretation that the soft X-ray excess is mainly due to inverse-Compton scattering. In fact, for each exposure, the fraction of accretion power dissipated within the warm corona is several times greater than the cooling due to the lamppost.

\section{Discussion and Conclusions}

We have reported on the first \textit{HST/XMM-Newton} monitoring campaign of the US-NLSy1 galaxy RBS\,1332. In the following, we will summarize and discuss our findings.\\

 \noindent\textbf{Variability properties:} In the observations the source showed significant flux variability of about 30\% on hourly timescales and up to a factor of 4 in about 10 days (see Fig.\ref{lc}). The amount of changes is compatible between the soft (0.5-2 keV) and the hard (2-10 keV) X-rays bands, as no significant spectral changes are witnessed during the exposures. Moreover, neither the UV nor the FUV data seem to vary during the observational campaign (see Fig.~\ref{omlc}). 

Given the compatible flux observed in the soft X-rays and ultraviolet bands (e.g. Table~\ref{components}), one could expect the variable soft X-rays to illuminate the outer disc and induce variability in the optical/UVs. In particular, the X-ray reprocessing would cause optical/UV reverberation lags as observed in long and richly sampled light curves of AGN \citep[e.g.][]{Mason2002,Arevalo2005,Alston2013,Lohfink2014,Edelson2015,Buisson2017,Edelson2019,Cackett2023}. The lack of variability we observe in the optical/UV bands can be explained by the short timescales investigated here, down to which X-rays variations are likely smeared in the outer disc this weakening any signal associated to reverberation \citep[e.g.][]{Smith2007,Robertson2015,Kammoun2021}. Any possible reverberated signal would, in fact, be diluted by the intrinsic UV emission. See \citet{Jin2017b} for a thorough discussion on this subject. 
\\

\noindent\textbf{Black hole mass estimates:}
The significant variability of the X-ray light curve can be used to estimate the supermassive black hole mass. Both long and short term X-ray variability was found to be anti-correlated with the AGN’s luminosity and black hole mass \citep[][]{Barr1986,Green1993,Lawrence1993,Markowitz2003,Papadakis2004,McHardy2006,Vagnetti2016,Paolillo2023}. Thus, we computed the light curves of RBS\,1332 for all the available observations, also including the archival (obs. ID.s 0741390201 and 0741390401) and extracted all the 2-10 keV light curves with a time bin of 500 s. Then, we calculated  in each 20 ks-long segment the normalized excess variance \citep[][]{Vaughan2003}, found to be $\sigma^{2}_{\rm nxs}=0.020\pm$0.009. Using the relations by \citet{Ponti2012}, \citealt[but see also][]{Tortosa2023} this quantity yields to a BH mass M$_{\rm BH}=(7.4\pm1.0) \times10^{6}M_{\odot}$. This estimate, that is compatible with previous values using the L$_{\rm 5100 \AA}$ and FWHM$_{\rm H\beta}$ \citep[see discussion in][]{Xu2021}, is about a factor of 2 smaller than the one inferred from the FUV analysis in Sect. 3.1 or with the single epoch estimate by \citet{Wu2022}. Considering an average value for the 2-10 keV luminosity and adopting the bolometric correction by \citet[][]{Duras2020}, we obtain a bolometric luminosity of L$_{\rm Bol}=8.4 \times10^{44}$ erg s$^{-1}$. This value, coupled with the BH mass derived from the normalized excess variance leads to an Eddington ratio of 90\%, about a factor of 2 larger than the one derived from analysis of the FUV spectrum.

\indent The discrepancy between the two values of Eddington ratio can be easily reconciled by the fact that objects with high accretion rates (super-Eddington accreting massive black holes, SEAMBHs) -- from significant Eddington fractions to the super-Eddington regime -- have been found to possess smaller broad-line region (BLR) sizes with respect to normal quasars \citep{Du2015}, probably related to altered disk and torus geometries. This in turn implies a higher BLR velocity dispersion, and thus a systematic overestimation of the $M_{\rm BH}$ value derived from single-epoch relations in the case of SEAMBHs. Hence, our difference in the $M_{\rm BH}$ values could be easily ascribed to accretion processes close to the Eddington limit powering the RBS\,1332 central engine and shaping its internal structure accordingly \citep[see also][]{Jin2017b}. We note that our results, with the exception of \textsc{AGNSED} do not depend on the BH mass and that, irrespectively of the value for the BH mass we may adopt, RBS\,1332 is compatible with being efficiently accreting matter.
\\

\noindent \textbf{Properties of the UV absorption features:} The main UV high-ionization transitions -- Ly$\alpha$, N {\footnotesize V} $\lambda$1241 and C {\footnotesize IV} $\lambda$1549 -- all show absorption features extended over a wide range of velocities, spanning $\sim$3200 km s$^{-1}$ from $\sim$--1500 km s$^{-1}$ to $\sim$1700 km s$^{-1}$. These features do not vary over a time span of $\sim$11.6 rest-frame days, implying electron densities $n_{\rm e} < 3.6 \times 10^5$ cm$^{-3}$ under the assumption of photoionization-driven absorption variability \citep{Barlow1992,Trevese2013}. Despite their moderate EWs (from $\sim$40 km s$^{-1}$ to $\sim$130 km s$^{-1}$, corresponding to minimum column densities of 2--12 $\times 10^{13}$ cm$^{-2}$), the presence of such troughs together with the evidence of blue-shifted emission-line wings associated with forbidden transitions \citep[see e.g.][]{Saturni2021} points at the existence of a mildy ionized outflow powered by the RBS\,1332 central engine, that can be further studied with dedicated investigations on its physical properties and geometry.

Such winds are ubiquitous in AGN at several redshifts \citep[e.g.,][]{Kakkad2016,Kakkad2018} and luminosities \citep[e.g.,][]{Vietri2018}, often extending to the whole host galaxy \citep[e.g.,][]{Perna2015,Perna2017} and thus playing an important role in regulating AGN activity and star formation by altering the amount of available gas for both these processes \citep[e.g.,][]{Cattaneo2009,Fabian2012}. Investigating with future dedicated studies the nature of the outflow at work in the RBS\,1332 environment will therefore be crucial to understand its impact on the host galaxy environment and evolution \citep[e.g.,][]{Cicone2018}.\\

\noindent \textbf{Spectral modeling and global picture:}  We tested four different models to account for the multi-epoch UV-to-X-ray data obtained in the context of our monitoring campaign: blurred relativistic reflection, warm Comptonization (also including \textsc{AGNSED}), and a model that combines both (relativistic reflection and warm-Comptonization), \textsc{reXcor}.
In the relativistic reflection scenario, the X-ray emission is in fact dominated by reflection, and the observed variability is mostly due to changes of the reflected flux. This could in turn be related to variations in the geometry of the disc-corona; for example, if the corona is a lamp-post source, a variation of the coronal height above the disc imply a variation of in the solid angle subtended by the disc.\\
\indent However, we find that the the broadband emission spectrum and the soft X-ray excess are statistically best modelled in the framework of the two-coronae model. A similar conclusion was drawn by \citet{Xu2021} using Swift and XMM-Newton archival exposures.\\
\indent We found the overall emission spectrum of RBS\,1332 emerges from three distinct components: i) a fairly constant outer disk ($\leq$200 R$_{\rm g}$) with T$_{\rm disk}\sim$1eV; ii) a patchy warm corona Compton up-scattering about $\sim$20\% of the underlying seed photons and extending from about 10 R$_{\rm g}$ to $\sim$200 R$_{\rm g}$); iii) a very compact ($\leq$10 R$_{\rm g}$) hot corona with a soft spectrum. 
The physical and geometrical parameters of the hot corona are moderately variable during the campaign, as indicated by the fits with \textsc{AGNSED} and \textsc{reXcor}. Also the radial extension of the warm corona shows a significant variability and is positively correlated with the accretion rate, while there is no clear trend with its temperature or optical depth. The same results were obtained by \cite{Palit2024arXiv} for a sample of Seyfert galaxies.
Despite the extreme soft X-ray excess observed in this source, its photon index, the temperature and the opacity are fairly standard when compared with results from other monitoring campaigns \citep[][]{Ursini2018,Middei2018,Porquet2018,Middei2019,Ursini2020,Middei2020}, archival studies \citep[][]{Petrucci2018,Petrucci2020,Xiang2022,Ballantyne2024,Palit2024arXiv} and distant quasars \citep[][]{Marinucci2022,Vaia2024arXiv}.\\
\indent The extension of the hot corona appears compact (R$_{\rm hot}\sim$ 10 R$_{g}$) and characterized by the ultra-soft observed hard power-law ($\Gamma\sim2.2$). This can be possibly explain by the interplay between a small heating power sustaining the hot corona and the efficient cooling due to the large fraction of seed photons from the warm corona. An other possibility is provided by the presence of a puffed-up disk-layer between the warm coronal region and the outer disk \citep[][]{Jin2017b}. This bloated layer of matter would be illuminated by the hard continuum, part of which could either pierce through it reaching the outer disk, or being reprocessed thus eventually producing a weak reflection signal.\\
\indent Finally, although limited by the moderate parameter range in the available \textsc{reXcor} tables, we found that the soft X-ray excess in RBS,1332 is primarily attributable to warm Comptonization with reprocessing of the primary continuum playing only a marginal role. This analysis, along with the findings by \citet{Jin2017} further supports the two-coronae model as a viable explanation for the multiwavelentgh emission spectrum of extreme AGN like the US-NLSy galaxies.

\begin{acknowledgements}
We thank the anonymous referee for their thorough reading of the manuscript. RM acknowledges Ioanna Psaradaki for insightful discussions on the UV spectra and acknowledges financial support from the ASI--INAF agreement n. 2022-14-HH.0. SB is an overseas researcher under the Postdoctoral Fellowship of Japan Society for the Promotion of Science (JSPS), supported by JSPS KAKENHI Grant Number JP23F23773.  This work relies on archival data, software or online services provided by the Space Science Data Center\,--\,ASI, and it is based on observations obtained with XMM-Newton, an ESA science mission with instruments and contributions directly funded by ESA Member States and NASA.POP and MC acknowledges financial support from the High Energy french National Programme (PNHE) of the National Center of Scientific research (CNRS) and from the french spatial agency (CNES). BDM acknowledges support via Ram\'on y Cajal Fellowship (RYC2018-025950-I), the Spanish MINECO grants PID2022-136828NB-C44 and PID2020-117252GB-I00, and the AGAUR/Generalitat de Catalunya grant SGR-386/2021.

\end{acknowledgements}

\thispagestyle{empty}
\bibliographystyle{aa}
\bibliography{RBS1332.bib}
%\appendix

%\section{The serendipitous discovery of a star}
%\label{xmmobs}

\end{document}